\RequirePackage{fix-cm}
\documentclass[smallextended]{svjour3}       
\smartqed  
\usepackage{graphicx,amssymb,amsfonts,mathrsfs,latexsym,amsmath,amssymb}
\journalname{XXX}
\begin{document}

\title{Complete classification for simple root cyclic codes over local rings $\mathbb{Z}_{p^s}[v]/\langle v^2-pv\rangle$
} \subtitle{}

\titlerunning{Complete classification for simple-root cyclic codes over $\mathbb{Z}_{p^s}[v]/\langle v^2-pv\rangle$}

\author{Yuan Cao\and Yonglin Cao$^\ast$
}


\institute{$^\ast$Yonglin Cao (corresponding author) \at
              School of Mathematics and Statistics, Shandong University of Technology, Zibo, Shandong 255091, China \\
              \email{ylcao@sdut.edu.cn}\\
           Yuan Cao \at
              School of Mathematics and Statistics, Shandong University of Technology, Zibo, Shandong 255091, China \\
              \email{yuancao@sdut.edu.cn}                 
          }

\date{Received: date / Accepted: date}

\maketitle

\begin{abstract}
Let $p$ be a prime integer,  $n,s\geq 2$
be integers satisfying ${\rm gcd}(p,n)=1$, and denote $R=\mathbb{Z}_{p^s}[v]/\langle v^2-pv\rangle$.
Then $R$ is a local non-principal ideal ring of $p^{2s}$ elements.
First, the structure of any cyclic code over $R$ of length $n$ and a complete classification of all these codes are presented. Then the cardinality of
each code and dual codes of these codes are given. Moreover, self-dual cyclic codes over $R$ of length $n$ are investigated. Finally, we list some optimal $2$-quasi-cyclic self-dual linear codes over $\mathbb{Z}_4$ of length $30$ and extremal $4$-quasi-cyclic self-dual binary linear $[60,30,12]$ codes  derived
from cyclic codes over $\mathbb{Z}_{4}[v]/\langle v^2+2v\rangle$ of length $15$.

\end{abstract}

\section{Introduction}
\label{intro}
 The catalyst for the study of codes over rings was the discovery of the connection
between the Kerdock and Preparata codes, which are non-linear binary
codes, and linear codes over $\mathbb{Z}_4$ (see [3] and [4]). Soon after this discovery,
codes over many different rings were studied. This led to many new
discoveries and concreted the study of codes over rings as an important
part of the coding theory discipline. Since $\mathbb{Z}_4$ is a chain ring, it was natural to expand the theory to focus on alphabets that are finite commutative chain rings (See [1], [2], [5], [7--10],
 [12], [14] and [16], for examples).

\par
   In 1999, Wood in [18] showed that
for certain reasons finite Frobenius rings are the most general class of
rings that should be used for alphabets of codes. Then self-dual codes over commutative
Frobenius rings were investigated in Dougherty et al. [11]. Especially,
in 2014,
  codes over an extension of $\mathbb{Z}_4$ were studied in Yildiz et al. [19] and
[20] where many good $\mathbb{Z}_4$-codes were obtained as images. The
ring in the mentioned works was described as $\mathbb{Z}_4[u]/\langle u^2\rangle=\mathbb{Z}_4+u\mathbb{Z}_4$ ($u^2=0$) which is a local non-principal
ring.

\par
    Let $A$ be an arbitrary finite commutative ring with identity $1\neq 0$, $A^{\times}$ the multiplicative group of units of
$A$ and $a\in
A$. We denote by $\langle a\rangle_A$, or $\langle a\rangle$ for
simplicity, the ideal of $A$ generated by $a$, i.e. $\langle
a\rangle_A=aA$. For any ideal $I$ of $A$, we will identify the
element $a+I$ of the residue class ring $A/I$ with $a$ (mod $I$) in this paper.

\par
   For any positive integer $N$, let
$A^N=\{(a_0,a_1,\ldots,a_{N-1})\mid a_i\in A, \ i=0,1,\ldots, N-1\}$ which is an $A$-module with componentwise addition and scalar multiplication by elements of $A$. Then an $A$-submodule ${\cal C}$ of $A^N$ is called a \textit{linear code} of length $N$ over $A$.
For any vectors $a=(a_0,a_1,\ldots,a_{N-1}), b=(b_0,b_1,\ldots,b_{N-1})\in A^N$.
The usual \textit{Euclidian inner product} of $a$ and $b$ is defined by
$[a,b]=\sum_{j=0}^{N-1}a_jb_j\in A$.
Then $[-,-]$ is a symmetric and non-degenerate bilinear form on the $A$-module
$A^N$. Let ${\cal C}$ be a linear code over $A$ of length $N$. The \textit{dual code}
of ${\cal C}$ is defined by ${\cal C}^{\bot}=\{a\in A^N\mid [a,b]=0, \ \forall
b\in {\cal C}\}$, and ${\cal C}$ is said to be \textit{self-dual} if ${\cal C}={\cal C}^{\bot}$.
We use the natural connection of cyclic codes
to polynomial rings, where $c=(c_0,c_1,\ldots, c_{N-1})\in A^N$ is viewed as
$c(x)=\sum_{j=0}^{N-1}c_jx^j$ and the cyclic code ${\cal C}$ is an ideal in the polynomial residue ring
$A[x]/\langle x^N-1\rangle$.

\par
  A linear code ${\cal C}$ of over $A$ length $N$ is said to be \textit{cyclic} if
 $(a_{N-1},a_0,a_1,\ldots$, $a_{N-2})\in {\cal C}$ for all
$(a_0,a_1,\ldots,a_{N-1})\in{\cal C}$. Let $A$ be a local ring with residue class field $F$. Then cyclic codes over $R$ of length $N$ are called
\textit{simple root cyclic codes} if ${\rm gcd}(N, {\rm char}(F))=1$.

\par
   In this paper, let $p$ be an arbitrary prime number, $s$ an integer satisfying $s\geq 2$. We adopt the following notations:

\par
 $\bullet$ $\mathbb{Z}_{p}=\mathbb{Z}/\langle p\rangle=\{0,1,2,\ldots,p-1\}$ in which the arithmetic is done modulo $p$. Then $\mathbb{Z}_{p}$ is a finite field.

\par
 $\bullet$ $\mathbb{Z}_{p^s}=\mathbb{Z}/\langle p^s\rangle=\{0,1,2,\ldots,p^s-1\}$ in which the arithmetic is done modulo $p^s$. Then $\mathbb{Z}_{p^s}$ is a finite chain ring.

\par
 $\bullet$ $R=\mathbb{Z}_{p^s}[v]/\langle v^2-pv\rangle=\{a+bv\mid a,b\in \mathbb{Z}_{p^s}\}=\mathbb{Z}_{p^s}+v\mathbb{Z}_{p^s}
\ (v^2=pv)$ in which the operations are defined by:
\begin{center}
$\alpha+\beta=(a+b)+v(c+d)$ and $\alpha\beta=ac+(ad+bc+pbd)v$.
\end{center}

\noindent
for any $\alpha=a+bv,\beta=c+dv\in \mathbb{Z}_{p^s}+v\mathbb{Z}_{p^s}$ with $a,b,c,d\in \mathbb{Z}_{p^s}$.
Then $R$ is a local Frobenius non-chain
ring of $p^{2s}$ elements.

\par
  For the special case of $p=s=2$, linear codes over $\mathbb{Z}_{4}+v\mathbb{Z}_{4}$ $(v^2=2v)$ were studied in [13]. In the paper, a duality preserving
Gray map was given and was used to present MacWilliams identities and self-dual
codes. Connections between these self-dual codes and real unimodular lattices
was also discussed. Some extremal Type II $\mathbb{Z}_{4}$-codes was provided as images of
codes over this ring. $\mathbb{Z}_{4}$-codes that was images of linear codes over the studied
ring are characterised through automorphism groups and some well-known
families of $\mathbb{Z}_{4}$-codes was
proved to be linear over it.

\par
   In this paper, let $n$ be a positive integer satisfying
${\rm gcd}(p,n)=1$. We consider
 the following questions
 for cyclic codes over $R=\mathbb{Z}_{p^s}+v\mathbb{Z}_{p^s}$ $(v^2=pv)$ of length $n$:

\par
  (Q-1) Present all distinct cyclic codes over $R$ by giving their generators precisely, and count the number of these codes explicitly.

\par
  (Q-2) For each code ${\cal C}$ presented above, determine the number of codewords contained in ${\cal C}$ and
give the dual code of ${\cal C}$ precisely.

\par
  (Q-3) Determine the self-duality for cyclic codes over $R$.

\vskip3mm \par
   The present paper is organized as follows.  In section 2, we sketch the basic theory of Galois rings and linear codes over Galois rings needed in this paper.
In Section 3, we give a canonical form decomposition for any cyclic code over $R$ of length $n$
and list all distinct codes by their generator sets. Using this decomposition, we
obtain the dual code of these codes and investigate their self-duality in Section 4. In Section 5, we focus our attention on cyclic code over $\mathbb{Z}_4+v\mathbb{Z}_4$. Especially, we present explicitly all $583443$ cyclic code over $\mathbb{Z}_4+v\mathbb{Z}_4$ of length $15$ and $315$ self-dual codes among them. Finally, we obtain $162$ good self-dual quasi-cyclic codes
over $\mathbb{Z}_4$ of index $2$ and length $30$ with minimum Lee weight $10$ and $12$. From these
$\mathbb{Z}_4$-codes, we derive $70$ quasi-cyclic binary extremal self-dual $[60,30,12]$ codes
of index $4$.



\section{Preliminaries}
\noindent
  In this section, we sketch the basic theory of Galois rings and linear codes over Galois rings rings needed in this paper.

\par
   In this paper, we will regard $\mathbb{Z}_p$ as a subset of the ring $\mathbb{Z}_{p^s}$ although
$\mathbb{Z}_p$ is not a subring of $\mathbb{Z}_{p^s}$. Then every element $a$ of $\mathbb{Z}_{p^s}$ has a
unique $p$-adic expansion:
$a=a_0+a_1p+\ldots+a_{s-1}p^{s-1}, \ a_0,a_1,\ldots,a_{s-1}\in \mathbb{Z}_p.$
Define $\overline{a}=a_0=a$ (mod $p$) for all $a\in \mathbb{Z}_{p^s}$. Then $^{-}$ is a surjective homomorphism
of rings from $\mathbb{Z}_{p^s}$ onto $\mathbb{Z}_{p}$ and can be extended to a surjective homomorphism
of rings from $\mathbb{Z}_{p^s}[x]$ onto $\mathbb{Z}_{p}[x]$ in the natural way:
$$\overline{f}(x)=\overline{f(x)}=\sum_{k}\overline{f}_kx^k, \ \forall f(x)=\sum_{k}f_kx^k\in \mathbb{Z}_{p^s}[x]
\ {\rm where} \ a_k\in \mathbb{Z}_{p^s}.$$

\par
  Let $f(x)$ be a monic polynomial in $\mathbb{Z}_{p^s}[x]$. Then $f(x)$ is said to be
\textit{basic irreducible} if $\overline{f}(x)$ is an irreducible polynomial in $\mathbb{Z}_{p}[x]$.
Now, let $f(x)$ be a fixed monic basic irreducible in $\mathbb{Z}_{p^s}[x]$ of degree $m\geq 1$ and set
$$A_f=\mathbb{Z}_{p^s}[x]/\langle f(x)\rangle=\{\sum_{i=0}^{m-1}a_ix^i\mid a_i\in \mathbb{Z}_{p^s}, \ i=0,1,\ldots,m-1\}$$
in which the arithmetic is done modulo $f(x)$. Then from Wan [17] Theorem 14.1, Lemma 4.4 and Theorem 14.8, we deduce the following conclusion.

\vskip 3mm\noindent
  {\bf Lemma 2.1} \textit{Using the notations above, $A_f$ is a Galois ring
of characteristic $p^s$ and cardinality $p^{sm}$.  Moreover,}

\par
  (i) \textit{Every element $\alpha$ of $A_f$ has a unique $p$-adic expansion}:
$$\alpha=t_0(x)+pt_1(x)+\ldots+p^{s-1}t_{s-1}(x), \ t_0(x),t_1(x),\ldots,t_{s-1}(x)\in T_f$$
\textit{where $T_f=\{\sum_{j=0}^{m-1}a_jx^j\mid a_j\in \mathbb{Z}_p\}$ is the Teichm\"{u}ller
set of $A_f$. Then $\alpha$ is invertible, i.e. $\alpha\in A_f^\times$, if and only if $t_0(x)\neq 0$.
Hence $|A_f^\times|=(p^m-1)p^{(s-1)m}$}.

\par
 (ii) \textit{$A_f$ is a finite chain ring with maximal ideal $\langle p\rangle=pA_f$,
$s$ is the nilpotency index of $p$ and all distinct ideals of $A_f$ are given by}:
$$\{0\}=\langle p^s\rangle\subset\langle p^{s-1}\rangle\subset\ldots\subset\langle p\rangle\subset\langle p^0\rangle=A_f,$$
\textit{where $\langle p^j\rangle=p^jA_f$ for all $j=0,1,\ldots,s$}.

\vskip 3mm \par
  For any $\alpha\in A_f$, by Lemma 2.1(i) there is a unique integer $k$, $0\leq k\leq s-1$, such that
$\alpha=p^k\alpha_0$ for some $\alpha_0\in A_f^\times$. We call $k$ the \textit{$p$-degree} of $\alpha$
and denote by $k=\|\alpha\|_p$. If $\alpha=0$ we write $\|\alpha\|_p=s$ for convenience.

\par
   Let $L$ be a positive integer and $A_f^L=\{(\alpha_1,\ldots,\alpha_L)\mid \alpha_1,\ldots,\alpha_L\in A_f\}$
the free $A_f$-module under
componentwise addition and multiplication with elements from $A_f$. Recall that a \textit{linear code} $C$ of length $L$
over the Galois ring $A_f$ is defined as an $A_f$-submodule of $A_f^L$, and $C$ is said to be \textit{nontrivial} if $C\neq A_f^L$ and $C\neq 0$.
Now, let $C$ be a linear code over $A_f$ of length $L$. By [14]
Definition 3.1, a matrix $G$ is called a \textit{generator matrix} for $C$ if the rows of $G$ span $C$ and none of them can be written as an $A_f$-linear combination of the other rows of $G$. Furthermore, a generator matrix $G$ is
said to be \textit{in standard form} if there is a suitable permutation matrix $U$ of size $L\times L$ such that
\begin{equation}
G=\left(\begin{array}{ccccc}
p^0I_{k_0} & M_{0,1}     & \ldots       & M_{0,s-1}       & M_{0,s}\cr
0            & p I_{k_1} & \ldots   & p M_{1,s-1}   &  p M_{1,s}\cr
\ldots            & \ldots           & \ldots & \ldots &  \ldots\cr
0            & 0           & 0             & p^{s-1} I_{k_{s-1}} & p^{s-1} M_{s-1,s} \end{array}\right)U
\end{equation}
where the columns are grouped into blocks of sizes $k_0,k_1,\ldots,k_{s-1}, k$
with $k_i\geq 0$ and $k=L-(k_0+k_1+\ldots+k_{s-1})$. Of course, if $k_i=0$, the matrices $p^iI_{k_i}$ and $p^iM_{i,j}$ ($i<j$) are suppressed in $G$.
From [14] Proposition 3.2 and Theorem 3.5, we deduce the following.

\vskip 3mm \noindent
   {\bf Lemma 2.2} \textit{Let $C$ be a nonzero linear code of length $L$ over $A_f$. Then $C$ has a generator matrix in standard form as in Equation $(1)$. In this case, the number of codewords in $C$ is equal to}
$$|C|=|T_f|^{sk_0+(s-1)k_1+(s-2)k_2+\ldots+2k_{s-2}+k_{s-1}}=p^{m\sum_{i=0}^{s-1}(s-i)k_i}.$$

\vskip 3mm\par
   In particular, all distinct nontrivial linear codes of length $2$  over
$A_f$ had been listed (cf. [6] Example 2.5). Moreover, we have

\vskip 3mm \noindent
   {\bf Theorem 2.3} \textit{Using the notations above, every linear code $C$ of length $2$ over
the Galois ring $A_f=\mathbb{Z}_{p^s}[x]/\langle f(x)\rangle$ satisfying the following condition
\begin{equation}
(0, a+pb)\in C, \ \forall (a,b)\in C
\end{equation}
has one and only one of the following matrices $G$ as their generator matrices}:

\vskip 2mm  \noindent
  $\bullet$ \textit{If $s=2$, there are $p^m+5$ matrices}:

\vskip 2mm \par
   (I) \textit{$G=(p w(x),1)$, $w(x)\in T_f$}.

\vskip 2mm \par
   (II) \textit{$G=(0,p)$}.

\vskip 2mm \par
   (III) \textit{$G=p^kI_2$ where $I_2$ is the identity matrix of order $2$, $0\leq k\leq 2$}.

\vskip 2mm \par
   (IV) \textit{$G=\left(\begin{array}{cc}0 & 1\cr
p & 0\end{array}\right)$}.

\vskip 2mm  \noindent
  $\bullet$ \textit{If $s=3$, there are $4p^m+7$ matrices}:

\vskip 2mm \par
   (I) \textit{$G=(p^{2}w(x),1)$ and $G=(-p+p^{2}w(x),1)$, where $w(x)\in T_f$}.

\vskip 2mm \par
   (II) \textit{$G=(0,p^2)$, and $G=(p^2w(x),p)$ where $w(x)\in T_f$}.

\vskip 2mm \par
   (III) \textit{$G=p^kI_2$, $0\leq k\leq 3$}.

\vskip 2mm \par
   (IV) \textit{$G=\left(\begin{array}{cc}0 & 1\cr
p & 0\end{array}\right)$, and $G=\left(\begin{array}{cc}p w(x) & 1\cr
p^2 & 0\end{array}\right)$ where $w(x)\in T_f$}.

\vskip 2mm \par
   (V) \textit{$G=\left(\begin{array}{cc}0 & p\cr
p^2 & 0\end{array}\right)$}.

\vskip 2mm  \noindent
  $\bullet$ \textit{If $s=4$, there are $9p^m+9$ matrices}:

\vskip 2mm \par
   (I) \textit{$G=(p^{3}w(x),1)$ and $G=(-p+p^{3}w(x),1)$, where $w(x)\in T_f$}.

\vskip 2mm \par
   (II) \textit{$G=(0, p^{3})$, $G=(p^{3}w(x),p^{2})$, $G=(p^{3}w(x),p)$ and $G=(-p^{2}+p^{3}w(x),p)$, where $w(x)\in T_f$}.

\vskip 2mm \par
   (III) \textit{$G=p^kI_2$, $0\leq k\leq 4$}.

\vskip 2mm \par
   (IV)  \textit{$G=\left(\begin{array}{cc}0 & 1\cr
p & 0\end{array}\right)$, $G=\left(\begin{array}{cc}p w(x) & 1\cr
p^2 & 0\end{array}\right)$,
$G=\left(\begin{array}{cc}-p+p^{2} w(x) & 1\cr
p^3 & 0\end{array}\right)$ and $G=\left(\begin{array}{cc}p^{2} w(x) & 1\cr
p^3 & 0\end{array}\right)$,  where $w(x)\in T_f$}.

\vskip 2mm \par
   (V) \textit{$G=\left(\begin{array}{cc}0 & p\cr
p^2 & 0\end{array}\right)$, $G=\left(\begin{array}{cc}0 & p^2\cr
p^3 & 0\end{array}\right)$, and $G=\left(\begin{array}{cc} p^2w(x) & p\cr
p^3 & 0\end{array}\right)$ where $w(x)\in T_f$}.

\vskip 2mm  \noindent
  $\bullet$ \textit{If $s\geq 5$, there are $(s-1)^2p^m+2s+1$ matrices}.

\vskip 2mm \par
   (I) \textit{$2p^m$ matrices: $G=(p^{s-1}w(x),1)$ and $G=(-p+p^{s-1}w(x),1)$, where $w(x)\in T_f$}.

\vskip 2mm \par
   (II) \textit{$(2s-5)p^m+1$ matrices: $G=(0, p^{s-1})$, $G=(p^{s-1}w(x),p^{s-2})$, $G=(p^{s-1}w(x),p^k)$ and $G=(-p^{k+1}+p^{s-1}w(x),p^k)$ with $1\leq k\leq s-3$, where $w(x)\in T_f$}.

\vskip 2mm \par
   (III) \textit{$s+1$ matrices:  $G=p^kI_2$, $0\leq k\leq s$}.

\vskip 2mm \par
   (IV) \textit{$(2s-5)p^m+1$ matrices: $G=\left(\begin{array}{cc}0 & 1\cr
p & 0\end{array}\right)$, $G=\left(\begin{array}{cc}p w(x) & 1\cr
p^2 & 0\end{array}\right)$,
$G=\left(\begin{array}{cc}-p+p^{t-1} w(x) & 1\cr
p^t & 0\end{array}\right)$ and $G=\left(\begin{array}{cc}p^{t-1} w(x) & 1\cr
p^t & 0\end{array}\right)$, where $3\leq t\leq s-1$ and $w(x)\in T_f$}.

\vskip 2mm \par
   (V) \textit{$(s-3)^2p^m+s-2$ matrices: $G=\left(\begin{array}{cc}0 & p^k\cr
p^{k+1} & 0\end{array}\right)$ with $1\leq k\leq s-2$, $G=\left(\begin{array}{cc} p^{k+1}w(x) & p^k\cr
p^{k+2} & 0\end{array}\right)$ with $1\leq k\leq s-3$, $G=\left(\begin{array}{cc} -p^{k+1}+p^{k+t-1}w(x) & p^k\cr
p^{k+t} & 0\end{array}\right)$ and $G=\left(\begin{array}{cc} p^{k+t-1}w(x) & p^k\cr
p^{k+t} & 0\end{array}\right)$ with $3\leq t\leq s-k-1$ and $1\leq k\leq s-4$, where $w(x)\in T_f$}.

\vskip 3mm \noindent
   {\bf Proof.} See Appendix.
\hfill $\Box$


\section{Structure and enumeration of cyclic codes over $R$ of length $n$}
In this section, we list all distinct cyclic codes of length $n$ over the ring $R=\mathbb{Z}_{p^s}+v\mathbb{Z}_{p^s}$ ($v^2=pv$),
i.e. all distinct ideals of the ring $R[x]/\langle x^{n}-1\rangle$, where $s\geq 2$ and ${\rm gcd}(p,n)=1$. From now on,
we denote
$$\mathcal{A}=\mathbb{Z}_{p^s}[x]/\langle x^n-1\rangle=\{\sum_{i=0}^{n-1}a_ix^i\mid a_i\in \mathbb{Z}_{p^s}, \ i=0,1,\ldots,n-1\}$$
in which the arithmetic is done modulo $x^n-1$. It is clear that $\mathcal{A}$ is a finite commutative ring containing $\mathbb{Z}_{p^s}$
as its subring, and set
$$\mathcal{A}+v\mathcal{A}=\mathcal{A}[v]/\langle v^2-pv\rangle=\{\alpha+\beta v\mid \alpha,\beta\in \mathcal{A}\} \
(v^2=pv)$$
in which $v$ is an indeterminate over $\mathcal{A}$ and the arithmetic on $\mathcal{A}+v\mathcal{A}$
is done modulo $v^2-pv$. Precisely, for any $\alpha_1,\alpha_2,\beta_1, \beta_2\in\mathcal{A}$ we have

\vskip 2mm\par
  $(\alpha_1+\beta_1 v)+(\alpha_2+\beta_2 v)=(\alpha_1+\alpha_2)+v(\beta_1+\beta_2)$,

\vskip 2mm\par
  $(\alpha_1+\beta_1 v)(\alpha_2+\beta_2 v)=\alpha_1\alpha_2+v(\alpha_1\beta_2+\beta_1\alpha_2+p\beta_1\beta_2)$.

\vskip 2mm\noindent
Then $\mathcal{A}+v\mathcal{A}$ is a finite commutative ring containing $\mathcal{A}$ as its subring.

\par
  Let $\alpha,\beta\in \mathcal{A}$. Then $\alpha$ and $\beta$ can be uniquely expressed as
$\alpha=\sum_{i=0}^{n-1}a_ix^i$ and $\beta=\sum_{i=0}^{n-1}b_ix^i$ respectively, where $a_i,b_i\in \mathbb{Z}_{p^s}$
for all $i=0,1,\ldots,n-1$. Now, we define a map $\Theta: \mathcal{A}+v\mathcal{A}\rightarrow R[x]/\langle x^{n}-1\rangle$ by
$$\Theta(\alpha+\beta v)=\sum_{i=0}^{n-1}\xi_ix^i, \
{\rm where} \ \xi_i=a_i+b_iv\in R, \ i=0,1,\ldots,n-1.$$
Then one can easily verify the following conclusion.

\vskip 3mm \noindent
  {\bf Theorem 3.1} \textit{The map $\Theta$ defined above is an isomorphism of rings from
$\mathcal{A}+v\mathcal{A}$ onto $R[x]/\langle x^{n}-1\rangle$}.

\vskip 3mm \par
  In the following, we will identify $\mathcal{A}+v\mathcal{A}$ with $R[x]/\langle x^{n}-1\rangle$
under the ring isomorphism $\Theta$. Therefore, in order to determine all cyclic codes
over $R$ of length $n$, we only need to determine all ideals of the ring $\mathcal{A}+v\mathcal{A}$. To do this, we
investigate the structure of the ring $\mathcal{A}$ first.

\par
  As ${\rm gcd}(p,n)=1$, by [17] Theorem 13.8 there are pairwise coprime monic basic irreducible polynomials
$f_0(x),f_1(x),\ldots,f_r(x)\in \mathbb{Z}_{p^s}[x]$ such that
\begin{equation}
x^n-1=f_0(x)f_1(x)\ldots f_r(x),
\end{equation}
where $\overline{f}_j(x)$ is irreducible in $\mathbb{Z}_{p}[x]$ and ${\rm deg}(f_j(x))=m_j$ for all $j=0,1,\ldots,r$. Especially, $f_0(x)=x-1$ with degree $m_0=1$.
For each integer $j$, $0\leq j\leq r$, denote $F_j(x)=\frac{x^n-1}{f_j(x)}\in \mathbb{Z}_{p^s}[x]$. By
${\rm gcd}(\overline{F}_j(x), \overline{f}_j(x))=1$, we see that $F_j(x)$ and $f_j(x)$ are coprime in $\mathbb{Z}_{p^s}[x]$
(cf. [17] Lemma 13.5). Hence there are polynomials $c_j(x), d_j(x)\in \mathbb{Z}_{p^s}[x]$ such that
\begin{equation}
c_j(x)F_j(x)+ d_j(x)f_j(x)=1.
\end{equation}
In this paper, we adopt the following notations:

\vskip 2mm\par
   $\bullet$ $e_j(x)\in \mathcal{A}$ satisfying
\begin{equation}
e_j(x)\equiv c_j(x)F_j(x)=1-d_j(x)f_j(x) \ ({\rm mod} \ x^n-1).
\end{equation}

\par
   $\bullet$ $\mathcal{K}_j=A_{f_j}=\mathbb{Z}_{p^s}[x]/\langle f_j(x)\rangle=\{\sum_{i=0}^{m_j-1}a_ix^i\mid a_i\in \mathbb{Z}_{p^s}, \ i=0,1,\ldots,m_j-1\}$
in which the arithmetic are done modulo $f_j(x)$.

\vskip 2mm\par
   $\bullet$ $\mathcal{K}_j+v\mathcal{K}_j=\mathcal{K}_j[v]/\langle v^2-pv\rangle=\{\alpha+\beta v\mid \alpha,\beta\in \mathcal{K}_j\}$
$(v^2=pv)$ in which $v$ is an indeterminate over $\mathcal{K}_j$ and the arithmetic on $\mathcal{K}_j+v\mathcal{K}_j$
are done modulo the monic polynomial $v^2-pv$ over $\mathcal{K}_j$.

\vskip 2mm \noindent
  Then $\mathcal{K}_j$ is a Galois ring of characteristic $p^s$ and cardinality $p^{sm_j}$ (cf. [17] Theorem 14.1),
$\mathcal{K}_j+v\mathcal{K}_j$ is a finite commutative ring containing $\mathcal{K}_j$ as its subring
and $|\mathcal{K}_j+v\mathcal{K}_j|=p^{2sm_j}$,
, for all
$j=0,1,\ldots,r$. From Wan [16] Theorem 2.7, one can easily deduce the following lemma.

\vskip 3mm \noindent
  {\bf Lemma 3.2} \textit{Using the notations above, we have the following conclusions}:

\par
  (i) \textit{$e_0(x)+e_1(x)+\ldots+e_r(x)=1$, $e_j(x)^2=e_j(x)$
and $e_j(x)e_l(x)=0$  in the ring $\mathcal{A}$ for all $0\leq j\neq l\leq r$}.

\par
  (ii) \textit{For each integer $j$, $0\leq j\leq r$, $\mathcal{A}e_j(x)$ is a subring
of $\mathcal{A}$ with $e_j(x)$ as its multiplicative identity. Define $\varphi_j(a(x))=a(x)e_j(x)$
$({\rm mod} \ x^n-1)$, $\forall a(x)\in \mathcal{K}_j$. Then $\varphi_j$ is
an isomorphism of rings from $\mathcal{K}_j$ onto $\mathcal{A}e_j(x)$ with inverse $\varphi_j^{-1}$}:
$$\varphi_j^{-1}(c(x))=c(x) \ ({\rm mod} \ f_j(x)), \ \forall c(x)\in \mathcal{A}e_j(x).$$

\par
  (iii) \textit{For any $a_j(x)\in \mathcal{K}_j$ and $j=0,1,\ldots,r$, define
$$
\varphi(a_0(x),a_1(x),\ldots,a_r(x))=\sum_{j=0}^r\varphi_j(a_j(x))
  =\sum_{j=0}^re_j(x)a_j(x) \ ({\rm mod} \ x^{n}-1).
$$
Then $\varphi$ is a ring isomorphism from $\mathcal{K}_0\times\mathcal{K}_1\times\ldots\times\mathcal{K}_r$ onto $\mathcal{A}$}.

\vskip 3mm \par
   We give a structural characterization for cyclic codes over $R$ of length $n$.

\vskip 3mm \noindent
  {\bf Theorem 3.3} \textit{Using the notations above, we have the following conclusions}.

\par
  (i) \textit{Define $\Phi(\xi_0,\xi_1,\ldots,\xi_r)=\sum_{j=0}^re_j(x)\xi_j$ $({\rm mod} \ x^n-1)$
$(\forall \xi_j\in \mathcal{K}_j+v\mathcal{K}_j, \ j=0,1,\ldots,r)$. Then
$\Phi$ is an isomorphism of rings from $(\mathcal{K}_0+v\mathcal{K}_0)\times(\mathcal{K}_1+v\mathcal{K}_1)\times
\ldots\times(\mathcal{K}_r+v\mathcal{K}_r)$ onto $\mathcal{A}+v\mathcal{A}$}.

\par
  (ii)  \textit{$\mathcal{C}$ is a cyclic code
 over $R$ of length $n$
if and only if for each integer $j$, $0\leq j\leq r$, there is a unique ideal $C_j$ of the ring $\mathcal{K}_j+v\mathcal{K}_j$
$(v^2=pv)$ such that}
$$\mathcal{C}=e_0(x)C_0\oplus e_1(x)C_1\oplus\ldots\oplus e_r(x)C_r \ ({\rm mod} \ x^n-1)$$
\textit{where $e_j(x)C_j=\{e_j(x)\alpha+v e_j(x)\beta\mid \alpha+\beta v\in C_j, \ \alpha,\beta \in \mathcal{K}_j\}
\subseteq  \mathcal{A}+v\mathcal{A}$
for all $j=0,1,\ldots,r$}. \textit{Moreover, the number of codewords in $\mathcal{C}$
is equal to $\prod_{j=0}^{r}|C_j|$}.

\vskip 3mm \noindent
   {\bf Proof} (i) Let $\xi=(\xi_0,\xi_1, \ldots, \xi_r)\in (\mathcal{K}_0+v\mathcal{K}_0)\times(\mathcal{K}_1+v\mathcal{K}_1)\times
\ldots\times(\mathcal{K}_r+v\mathcal{K}_r)$
where $\xi_j=\alpha_j+\beta_j v$ and $\alpha_j,\beta_j\in \mathcal{K}_j$ for all $0\leq j\leq r$.
By the definition $\varphi$ defined in Lemma 3.2(iii), we have
\begin{eqnarray*}
\Phi(\xi) &=&\sum_{j=0}^re_j(x)\xi_j
 =\sum_{j=0}^re_j(x)\left(\alpha_j+v\beta_j \right)
 =\sum_{j=0}^re_j(x)\alpha_j+v\sum_{j=0}^re_j(x)\beta_j\\
 &=&\varphi(\alpha_0,\alpha_1,\ldots,\alpha_r)+v\varphi(\beta_0,\beta_1,\ldots,\beta_r).
\end{eqnarray*}
Then for any $\eta=(\eta_0,\eta_1,\ldots,\eta_r)$, where $\eta_j=\gamma_j+v\delta_j$ with
$\gamma_j,\delta_j\in \mathcal{K}_j$ for all $j$, by $\Phi(\eta)=\varphi(\gamma_0,\gamma_1,\ldots,\gamma_r)
+v\varphi(\delta_0,\delta_1,\ldots,\delta_r)$ and $v^2=pv$ it follows that
\begin{eqnarray*}
 &&\Phi(\xi\eta) \\
 &=&\Phi(\xi_0\eta_0,\xi_1\eta_1,\ldots,\xi_r\eta_r)\\
 &=&\Phi(\alpha_0\gamma_0+v(\alpha_0\delta_0+\beta_0\gamma_0+p\beta_0\delta_0),\ldots,
    \alpha_r\gamma_r+v(\alpha_r\delta_r+\beta_r\gamma_r+p\beta_r\delta_r))\\
 &=&\varphi(\alpha_0\gamma_0,\ldots,\alpha_r\gamma_r)
   +v\varphi(\alpha_0\delta_0+\beta_0\gamma_0+p\beta_0\delta_0,
   \ldots,\alpha_r\delta_r+\beta_r\gamma_r+p\beta_r\delta_r)\\
 &=&\varphi(\alpha_0,\ldots,\alpha_r)\varphi(\gamma_0,\ldots,\gamma_r)  +v\left(\varphi(\alpha_0,\ldots,\alpha_r)\varphi(\delta_0,\ldots,\delta_r)\right.\\
 &&+\varphi(\beta_0,\ldots,\beta_r)\varphi(\gamma_0,\ldots,\gamma_r)
 +\left. p\varphi(\beta_0,\ldots,\beta_r)\varphi(\delta_0,\ldots,\delta_r)\right)\\
 &=&\Phi(\xi)\cdot \Phi(\eta),
\end{eqnarray*}
since $\varphi$ is a ring isomorphism from  $\mathcal{K}_0\times\mathcal{K}_1\times\ldots\times\mathcal{K}_r$ onto $\mathcal{A}$.
Similarly, we have $\Phi(\xi+\eta)=\Phi(\xi)+\Phi(\eta)$. Hence $\Phi$ is a ring isomorphism $\Phi$ from
$(\mathcal{K}_0+v\mathcal{K}_0)\times(\mathcal{K}_1+v\mathcal{K}_1)\times\ldots\times(\mathcal{K}_r+v\mathcal{K}_r)$ onto $\mathcal{A}+v\mathcal{A}$.

\par
  (ii) From the properties of ring isomorphisms
and direct product rings,  by (i) we conclude that $\mathcal{C}$ is a cyclic code over $R$ of length $n$, i.e. $\mathcal{C}$ is an ideal of $\mathcal{A}+v\mathcal{A}$,
if and only if for each integer $j$, $0\leq j\leq r$, there is a unique ideal $C_j$ of the ring $\mathcal{K}_j+v\mathcal{K}_j$
such that
\begin{eqnarray*}
\mathcal{C}&=&\Phi(C_0\times C_1\times\ldots\times C_r)
  =\{\Phi(\xi_0,\xi_1, \ldots, \xi_r)\mid \xi_j\in C_j, \ j=0,1,\ldots,r\}\\
 &=&\{\sum_{j=0}^re_j(x)\xi_j\mid \xi_j\in C_j, \ j=0,1,\ldots,r\}
 =\sum_{j=0}^re_j(x)\{\xi_j\mid \xi_j\in C_j\}.
\end{eqnarray*}
Hence $\mathcal{C}=\bigoplus_{j=0}^re_j(x)C_j$ (mod $x^n-1$)
and $|\mathcal{C}|=|C_0\times C_1\times\ldots\times C_r|=\prod_{j=0}^r|C_j|$.
\hfill $\Box$

\vskip 3mm\par
   Using the notations of Theorem 3.3, ${\cal C}=\bigoplus_{j=0}^re_j(x)C_j$ is called the \textit{canonical form decomposition} of the cyclic code ${\cal C}$ over $R$ of length $n$.

\par
   As $\mathcal{K}_j$ is a subring of $\mathcal{K}_j+v\mathcal{K}_j$, $\mathcal{K}_j+v\mathcal{K}_j$ can be regarded
as a $\mathcal{K}_j$-module. Precisely, $\mathcal{K}_j+v\mathcal{K}_j$ is a free $\mathcal{K}_j$-module with basis $\{1,v\}$.
Let $\mathcal{K}_j^2=\{(\alpha,\beta)\mid \alpha,\beta\in \mathcal{K}_j\}$. Then $\mathcal{K}_j^2$ is a  free $\mathcal{K}_j$-module
of rank $2$ with the componentwise addition and scalar multiplication. Now, define
$$\sigma: \mathcal{K}_j^2\rightarrow \mathcal{K}_j+v\mathcal{K}_j
\ {\rm via} \ (\alpha,\beta)\mapsto \alpha+\beta v \ (\forall \alpha,\beta\in \mathcal{K}_j).$$
Then it is obvious that $\sigma$ is an isomorphism of $\mathcal{K}_j$-modules from
$\mathcal{K}_j^2$ onto $\mathcal{K}_j+v\mathcal{K}_j$. Moreover, we have
the following key conclusion.

\vskip 3mm \noindent
  {\bf Lemma 3.4} \textit{Let $0\leq j\leq r$. Then $C_j$ is an ideal of the ring $\mathcal{K}_j+v\mathcal{K}_j$
if and only if there is a unique $\mathcal{K}_j$-submodule $S_j$ of $\mathcal{K}_j^2$ satisfying the following
condition}
\begin{equation}
(0,\alpha+p\beta)\in S_j, \ \forall (\alpha,\beta)\in S_j
\end{equation}
\textit{such that $\sigma(S_j)=C_j$}.

\vskip 3mm \noindent
   {\bf Proof} Let $C$ be an ideal
of $\mathcal{K}_j+v\mathcal{K}_j$. Then $C$ is a
$\mathcal{K}_j$-submodule of $\mathcal{K}_j+v\mathcal{K}_j$ satisfying
$v\xi\in C$ for any $\xi\in C$. Now, let
$S=\{(\alpha,\beta)\mid \alpha+v\beta\in C\}=\sigma^{-1}(C)$. Then it is obvious that
$S$ is a $\mathcal{K}_j$-submodule of $\mathcal{K}_j^2$ satisfying $C=\sigma(S)$. Moreover,
for any $(\alpha,\beta)\in S$, i.e. $\alpha+v\beta\in C$, by $v^2=pv$ in $\mathcal{K}_j+v\mathcal{K}_j$ it follows
that
$v(\alpha+p\beta)=v(\alpha+v\beta)\in C$. This implies
 $(0,\alpha+p\beta)\in S$.

\par
  Conversely, let $C=\sigma(S)$ and $S$ be a $\mathcal{K}_j$-submodule of $\mathcal{K}_j^2$ satisfying
Condition (6). Let $\alpha+v\beta\in C$ where $(\alpha,\beta)\in S$. For any $\xi+v\eta\in \mathcal{K}_j+v\mathcal{K}_j$ by
$v^2=pv$ and
$(0,\alpha+p\beta)\in S$ it follows that
\begin{eqnarray*}
\sigma^{-1}((\xi+v\eta)(\alpha+v\beta))&=&\sigma^{-1}\left(\xi(\alpha+v\beta)+\eta(v(\alpha+p\beta))\right)\\
  &=&\xi\sigma^{-1}(\alpha+v\beta)+\eta\sigma^{-1}(v(\alpha+p\beta))\\
  &=&\xi(\alpha,\beta)+\eta(0,\alpha+p\beta)\in S.
\end{eqnarray*}
This implies $(\xi+v\eta)(\alpha+v\beta)\in C$. Hence $C$ is an ideal of $\mathcal{K}_j+v\mathcal{K}_j$.
\hfill $\Box$

\vskip 3mm \par
   For any ideal $C$ of $\mathcal{K}_j+v\mathcal{K}_j$, its
\textit{annihilating ideal} is defined by
${\rm Ann}(C)=\{\beta\in \mathcal{K}_j+v\mathcal{K}_j\mid \alpha\beta=0, \ \forall \alpha\in C\}$. Now, by Lemma 2.2, Theorem 2.3 and Lemma 3.4 we can list all distinct ideals and their annihilating ideals of the ring $\mathcal{K}_j+v\mathcal{K}_j$ by the following theorem.

\vskip 3mm \noindent
  {\bf Theorem 3.5} \textit{Let $0\leq j\leq r$ and denote
$T_{f_j}=\{\sum_{i=0}^{m_j-1}a_ix^i\mid a_i\in\mathbb{Z}_p, \ i=0,1,\ldots, m_j-1\}.$
Then all distinct ideals $C_j$ and their annihilating ideals of the ring $\mathcal{K}_j+v\mathcal{K}_j$ are given by the following cases}.

\vskip 2mm  \noindent
  $\bullet$ \textit{If $s=2$, there are $p^{m_j}+5$ ideals $C_j$ given by the following table}:
\begin{center}
\begin{tabular}{lll|l}\hline
 N     &     $C_j$  & $|C_j|$  & ${\rm Ann}(C_j)$  \\ \hline
$p^{m_j}$ &  $\langle p w(x)+v\rangle$ & $p^{2m_j}$  & $\langle p(-1-w(x))+v\rangle$ \\
$1$ &  $\langle pv\rangle$ & $p^{m_j}$  & $\langle p,v\rangle$ \\
$3$ &  $\langle p^k\rangle$, $0\leq k\leq 2$ & $p^{(4-2k)m_j}$  & $\langle p^{2-k}\rangle$ \\
$1$ &  $\langle p,v\rangle$ & $p^{3m_j}$  & $\langle pv\rangle$ \\
\hline
\end{tabular}
\end{center}
\textit{where $w(x)\in T_{f_j}$ and $N$ is the number of ideals in the same row}.

\vskip 2mm  \noindent
  $\bullet$ \textit{If $s=3$, there are $4p^{m_j}+7$ ideals $C_j$ given by the following table}:
\begin{center}
\begin{tabular}{lll|l}\hline
 N     &     $C_j$  & $|C_j|$  & ${\rm Ann}(C_j)$  \\ \hline
$p^{m_j}$ &  $\langle p^2 w(x)+v\rangle$ & $p^{3m_j}$  & $\langle -p-p^2w(x)+v\rangle$ \\
$p^{m_j}$ &  $\langle -p+p^2 w(x)+v\rangle$ & $p^{3m_j}$  & $\langle -p^2w(x)+v\rangle$ \\
$1$ &  $\langle p^2v\rangle$ & $p^{m_j}$  & $\langle p,v\rangle$ \\
$p^{m_j}$ &  $\langle p^2w(x)+pv\rangle$ & $p^{2m_j}$  & $\langle p^2, p(-1-w(x))+v\rangle$ \\
$4$ &  $\langle p^k\rangle$, $0\leq k\leq 3$ & $p^{(6-2k)m_j}$  & $\langle p^{3-k}\rangle$ \\
$1$ &  $\langle p,v\rangle$ & $p^{5m_j}$  & $\langle p^2v\rangle$ \\
$p^{m_j}$ &  $\langle p^2,pw(x)+v\rangle$ & $p^{4m_j}$  & $\langle p^2(-1-w(x))+pv\rangle$ \\
$1$ &  $\langle p^2,pv\rangle$ & $p^{3m_j}$  & $\langle p^2,pv\rangle$ \\
\hline
\end{tabular}
\end{center}
\textit{where $w(x)\in T_{f_j}$ and $N$ is the number of ideals in the same row}.

\vskip 2mm  \noindent
  $\bullet$ \textit{If $s=4$, there are $9p^{m_j}+9$ ideals $C_j$ given by the following table}:
\begin{center}
\begin{tabular}{lll|l}\hline
 N     &     $C_j$  & $|C_j|$  & ${\rm Ann}(C_j)$  \\ \hline
$p^{m_j}$ &  $\langle p^3 w(x)+v\rangle$ & $p^{4m_j}$  & $\langle -p-p^3w(x)+v\rangle$ \\
$p^{m_j}$ &  $\langle -p+p^3 w(x)+v\rangle$ & $p^{4m_j}$  & $\langle -p^3w(x)+v\rangle$ \\
$1$ &  $\langle p^3v\rangle$ & $p^{m_j}$  & $\langle p,v\rangle$ \\
$p^{m_j}$ &  $\langle p^3w(x)+p^2v\rangle$ & $p^{2m_j}$  & $\langle p^2, p(-1-w(x))+v\rangle$ \\
$p^{m_j}$ &  $\langle p^3w(x)+pv\rangle$ & $p^{3m_j}$  & $\langle p^3, -p-p^2w(x)+v\rangle$ \\
$p^{m_j}$ &  $\langle -p^2+p^3w(x)+pv\rangle$ & $p^{3m_j}$  & $\langle p^3, -p^2w(x)+v\rangle$ \\
$5$ &  $\langle p^k\rangle$, $0\leq k\leq 4$ & $p^{(8-2k)m_j}$  & $\langle p^{4-k}\rangle$ \\
$1$ &  $\langle p,v\rangle$ & $p^{7m_j}$  & $\langle p^3v\rangle$ \\
$p^{m_j}$ &  $\langle p^2,pw(x)+v\rangle$ & $p^{6m_j}$  & $\langle p^3(-1-w(x))+p^2v\rangle$ \\
$p^{m_j}$ &  $\langle p^3,p^2w(x)+v\rangle$ & $p^{5m_j}$  & $\langle -p^2-p^3w(x)+pv\rangle$ \\
$p^{m_j}$ &  $\langle p^3,-p+p^2w(x)+v\rangle$ & $p^{5m_j}$  & $\langle -p^3w(x)+pv\rangle$ \\
$1$ &  $\langle p^2,pv\rangle$ & $p^{5m_j}$  & $\langle p^3,p^2v\rangle$ \\
$1$ &  $\langle p^3,p^2v\rangle$ & $p^{3m_j}$  & $\langle p^2,pv\rangle$ \\
$p^{m_j}$ &  $\langle p^3,p^2w(x)+pv\rangle$ & $p^{4m_j}$  & $\langle p^3,p^2(-1-w(x))+pv\rangle$ \\
\hline
\end{tabular}
\end{center}
\textit{where $w(x)\in T_{f_j}$ and $N$ is the number of ideals in the same row}.

\vskip 2mm  \noindent
  $\bullet$ \textit{If $s\geq 5$, there are $(s-1)^2p^{m_j}+2s+1$ ideals $C_j$ given by the following}.

\vskip 2mm \par
   (i) \textit{$p^{m_j}$ ideals: $C_j=\langle p^{s-1} w(x)+v\rangle$ with $|C_j|=p^{sm_j}$ and
${\rm Ann}(C_j)=\langle -p-p^{s-1}w(x)+v\rangle$, where $w(x)\in T_{f_j}$};

\vskip 2mm \par
   (ii) \textit{$p^{m_j}$ ideals: $C_j=\langle -p+p^{s-1} w(x)+v\rangle$ with $|C_j|=p^{sm_j}$ and
${\rm Ann}(C_j)=\langle -p^{s-1}w(x)+v\rangle$, where $w(x)\in T_{f_j}$};

\vskip 2mm \par
   (iii) \textit{$1$ ideal: $C_j=\langle p^{s-1}v\rangle$ with $|C_j|=p^{m_j}$ and
${\rm Ann}(C_j)=\langle  p,v\rangle$};

\vskip 2mm \par
   (iv) \textit{$p^{m_j}$ ideals: $C_j=\langle p^{s-1}w(x)+p^{s-2}v\rangle$ with $|C_j|=p^{2m_j}$ and
${\rm Ann}(C_j)=\langle  p^2, p(-1-w(x))+v\rangle$, where $w(x)\in T_{f_j}$};

\vskip 2mm \par
   (v) \textit{$(s-3)p^{m_j}$ ideals: $C_j=\langle p^{s-1}w(x)+p^kv\rangle$ with $|C_j|=p^{(s-k)m_j}$ and
${\rm Ann}(C_j)=\langle  p^{s-k}, -p-p^{s-k-1}w(x)+v\rangle$, where $w(x)\in T_{f_j}$ and $1\leq k\leq s-3$};

\vskip 2mm \par
   (vi) \textit{$(s-3)p^{m_j}$ ideals: $C_j=\langle -p^{k+1}+p^{s-1}w(x)+p^kv\rangle$ with $|C_j|=p^{(s-k)m_j}$ and
${\rm Ann}(C_j)=\langle  p^{s-k}, -p^{s-k-1}w(x))+v\rangle$, where $w(x)\in T_{f_j}$ and $1\leq k\leq s-3$};

\vskip 2mm \par
   (vii) \textit{$s+1$ ideals: $C_j=\langle p^k\rangle$ with $|C_j|=p^{2(s-k)m_j}$ and
${\rm Ann}(C_j)=\langle  p^{s-k}\rangle$};

\vskip 2mm \par
   (viii) \textit{$1$ ideals: $C_j=\langle p,v\rangle$ with $|C_j|=p^{(2s-1)m_j}$ and
${\rm Ann}(C_j)=\langle  p^{s-1}v\rangle$};

\vskip 2mm \par
   (ix) \textit{$p^{m_j}$ ideals: $C_j=\langle p^2,pw(x)+v\rangle$ with $|C_j|=p^{(2s-2)m_j}$ and
${\rm Ann}(C_j)=\langle  p^{s-1}(-1-w(x))+p^{s-2}v\rangle$, where $w(x)\in T_{f_j}$};

\vskip 2mm \par
   (x) \textit{$(s-3)p^{m_j}$ ideals: $C_j=\langle p^t,p^{t-1}w(x)+v\rangle$ with $|C_j|=p^{(2s-t)m_j}$ and
${\rm Ann}(C_j)=\langle  -p^{s-t+1}-p^{s-1}w(x)+p^{s-t}v\rangle$, where $w(x)\in T_{f_j}$ and $3\leq t\leq s-1$};

\vskip 2mm \par
   (xi) \textit{$(s-3)p^{m_j}$ ideals: $C_j=\langle p^t,-p+p^{t-1}w(x)+v\rangle$ with $|C_j|=p^{(2s-t)m_j}$ and
${\rm Ann}(C_j)=\langle  -p^{s-1}w(x)+p^{s-t}v\rangle$, where $w(x)\in T_{f_j}$ and $3\leq t\leq s-1$};

\vskip 2mm \par
   (xii) \textit{$s-2$ ideals: $C_j=\langle p^{k+1},p^kv\rangle$ with $|C_j|=p^{(2s-2k-1)m_j}$ and
${\rm Ann}(C_j)=\langle  p^{s-k},p^{s-k-1}v\rangle$, where $1\leq k\leq s-2$};

\vskip 2mm \par
   (xiii) \textit{$(s-3)p^{m_j}$ ideals: $C_j=\langle p^{k+2},p^{k+1}w(x)+p^kv\rangle$, $|C_j|=p^{(2s-2k-2)m_j}$ and
${\rm Ann}(C_j)=\langle  p^{s-k},p^{s-k-1}(-1-w(x))+p^{s-k-2}v\rangle$, where $w(x)\in T_{f_j}$ and $1\leq k\leq s-3$};

\vskip 2mm \par
   (xiv) \textit{$\frac{(s-3)(s-4)}{2}p^{m_j}$ ideals: $C_j=\langle p^{k+t},p^{k+t-1}w(x)+p^kv\rangle$ with $|C_j|=p^{(2s-2k-t)m_j}$ and
${\rm Ann}(C_j)=\langle  p^{s-k},-p^{s-k-t+1}-p^{s-k-1}w(x)+p^{s-k-t}v\rangle$, where $w(x)\in T_{f_j}$, $3\leq t\leq s-k-1$ and $1\leq k\leq s-4$};

\vskip 2mm \par
   (xv) \textit{$\frac{(s-3)(s-4)}{2}p^{m_j}$ ideals: $C_j=\langle p^{k+t},-p^{k+1}+p^{k+t-1}w(x)+p^kv\rangle$ with $|C_j|=p^{(2s-2k-t)m_j}$ and
${\rm Ann}(C_j)=\langle p^{s-k},-p^{s-k-1}w(x)+p^{s-k-t}v\rangle$, where $w(x)\in T_{f_j}$, $3\leq t\leq s-k-1$ and $1\leq k\leq s-4$}.

\vskip 3mm \noindent
   {\bf Proof} Let $s\geq 5$ and $C_j$ be an ideal of $\mathcal{K}_j+v\mathcal{K}_j$. Then
by Theorem 2.3 and Lemma 3.4, we have that
$C_j=\sigma(S)$ where $S$ is a linear code of length $2$ over the Galois ring
$\mathcal{K}_j=A_{f_j}=\mathbb{Z}_{p^s}[x]/\langle f_j(x)\rangle$ generated by one of
the $(s-1)^2p^{m_j}+2s+1$ matrices listed in Theorem 2.3 (by replacing $T_f$ with $T_{f_j}$).

\par
  (xiv) Let $S$ be generated by $G=\left(\begin{array}{cc} p^{k+t-1}w(x) & p^k\cr
p^{k+t} & 0\end{array}\right)$. Then by Lemma 2.2 and $|T_{f_j}|=p^{m_j}$ it follows
that $|S|=p^{m_j((s-k)+(s-(k+t)))}=p^{(2s-2k-t)m_j}$. Since $\sigma$ is a $\mathcal{K}_j$-module
isomorphism from $\mathcal{K}_j^2$ onto $\mathcal{K}_j+v\mathcal{K}_j$, we have
$C_j=\langle \sigma(p^{k+t}, 0), \sigma(p^{k+t-1}w(x), p^k)\rangle=\langle p^{k+t},p^{k+t-1}w(x)+ p^kv\rangle$ and $|C_j|=|S|$.

\par
  The other cases can be verified similarly. We omit these proofs here.
\hfill $\Box$

 \vskip 3mm \noindent
   {\bf Remark} When $j=0$, we have $f_0(x)=x-1$ with $m_0=1$. This implies $\mathcal{K}_0=\mathbb{Z}_{p^s}[x]/\langle x-1\rangle=\mathbb{Z}_{p^s}$ and $T_{f_0}=\mathbb{Z}_{p}$.
Hence $\mathcal{K}_0+v\mathcal{K}_0=\mathbb{Z}_{p^s}+v\mathbb{Z}_{p^s}$ ($v^2=pv$).
Then by replacing $T_{f_j}$ and $m_j$ with $\mathbb{Z}_{p}$ and $1$ respectively, all distinct ideals of the local ring $\mathbb{Z}_{p^s}+v\mathbb{Z}_{p^s}$ are listed in Theorem 3.5.

 \vskip 3mm \par
   Finally, from Theorems 3.3 and 3.5 we deduce the following corollary.

\vskip 3mm \noindent
  {\bf Corollary 3.6} \textit{Every cyclic code of length $n$ over $R=\mathbb{Z}_{p^s}+v\mathbb{Z}_{p^s}$  $(v^2=pv)$ can be constructed by the following two steps}:

\par
  (i) \textit{For each $j=0,1,\ldots,r$, choose an ideal $C_j$ of
${\cal K}_j+v{\cal K}_j$ listed in Theorem 3.5}.

\par
  (ii) \textit{Set ${\cal C}=\bigoplus_{j=0}^re_j(x)C_j$} (mod $x^{n}-1$).

\noindent
\textit{The number of codewords in ${\cal C}$ is
equal to $|{\cal C}|=\prod_{j=0}^r|C_j|$ and the minimum Hamming distance of ${\cal C}$ satisfies}
$d_{{\rm min}}({\cal C})\leq {\rm min}\{d_{{\rm min}}(e_j(x)C_j)\mid j=0,1,\ldots,r \}.$

\par
   \textit{Moreover, the number of cyclic
codes over $R$ of length $n$ is given by the formula}:
$\prod_{j=0}^r\left((s-1)^2p^{m_j}+2s+1\right).$



\section{Dual codes of cyclic codes over $\mathbb{Z}_{p^s}+v\mathbb{Z}_{p^s}$}
  In this section, we give the dual code of each cyclic code over the ring $R=\mathbb{Z}_{p^s}+v\mathbb{Z}_{p^s}$ ($v^2=pv$)
of length $n$ and investigate the self-duality of these codes.

\par
   For any $\alpha(x)=\sum_{i=0}^{n-1}\alpha_ix^i\in R[x]/\langle x^n-1\rangle$,
where $\alpha_i=a_i+vb_i\in R$ with $a_i,b_i\in\mathbb{Z}_{p^s}$ for all $0\leq i\leq n-1$,
we have $\alpha(x)=a(x)+vb(x)$ where $a(x)=\sum_{i=0}^{n-1}a_ix^i, b(x)=\sum_{i=0}^{n-1}b_ix^i
\in \mathcal{A}=\mathbb{Z}_{p^s}[x]/\langle x^n-1\rangle$. Now, we define
$$\tau(\alpha(x))=\alpha(x^{-1})=\alpha_0+\sum_{i=1}^{n-1}\alpha_ix^{n-i}
  =a(x^{-1})+vb(x^{-1}),$$
where $a(x^{-1})=a_0+\sum_{i=1}^{n-1}a_ix^{n-i}, b(x^{-1})=b_0+\sum_{i=1}^{n-1}b_ix^{n-i}\in
\mathcal{A}$. It can be verified easily that $\tau$ is a ring automorphism of $R[x]/\langle x^n-1\rangle$.
Moreover, it  is well known that

\vskip 3mm
\noindent
  {\bf Lemma 4.1} \textit{For any cyclic code $\mathcal{C}$ over $R$ of length $n$,
its dual code is given by
$\mathcal{C}^{\bot}=\tau({\rm Ann}(\mathcal{C}))
=\{a(x^{-1})\mid a(x)\in \mathcal{C}\},$
where ${\rm Ann}(\mathcal{C})=\{\eta\in R[x]/\langle x^n-1\rangle\mid \xi\eta=0,
\ \forall \xi\in \mathcal{C}\}$ is the annihilating ideal of $\mathcal{C}$ in $R[x]/\langle x^n-1\rangle$}.

\vskip 3mm
\par
   Since $R[x]/\langle x^n-1\rangle=\mathcal{A}+v\mathcal{A}$ $(v^2=pv)$ by Theorem 3.1, we see that the restriction
of $\tau$ on its subring $\mathcal{A}$ is a ring automorphism of $\mathcal{A}$. We still denote this
ring automorphism by $\tau$. Then $\tau(a(x))=a(x^{-1})$
for all $a(x)\in \mathcal{A}$.

\par
   Now, using the notations of Theorems 3.3 and 3.5, we present
the Euclidian dual code for any cyclic code over $R$ of length $n$.

 \vskip 3mm
\noindent
  {\bf Theorem 4.2} \textit{Let $\mathcal{C}$ be a cyclic code over $R$ of length $n$
with the canonical form decomposition ${\cal C}=\bigoplus_{j=0}^re_j(x)C_j$, where
$C_j$ is an ideal of $\mathcal{K}_j+v\mathcal{K}_j$ listed by Theorem 3.5 for $j=0,1,\ldots,r$.
For any integer $j$, $0\leq j\leq r$, we denote}
$\tau({\rm Ann}(C_j))=\{a(x^{-1})+vb(x^{-1})\mid a(x),b(x)\in\mathcal{K}_j, \
a(x)+vb(x)\in {\rm Ann}(C_j)\},$
\textit{where $x^{-1}=x^{n-1}$. Then
$$\mathcal{C}^{\bot}=\bigoplus_{j=0}^re_j(x^{-1})\tau({\rm Ann}(C_j)) \ ({\rm mod} \ x^n-1)$$
which is a cyclic code over $R$ of length $n$ as well}.

\vskip 3mm
\noindent
  {\bf Proof.} Let $\alpha,\beta\in R[x]/\langle x^n-1\rangle=\mathcal{A}+v\mathcal{A}$. By
Theorem 3.3(i), we have
$\alpha=\sum_{j=0}^re_j(x)\xi_j$ and $\beta=\sum_{j=0}^re_j(x)\eta_j$ where
$\xi_j,\eta_j\in \mathcal{K}_j+v\mathcal{K}_j$. Since $e(x)_j^2=e(x)_j$ and
$e_j(x)e_l(x)=0$ for all $0\leq j\neq l\leq r$ by Lemma 3.2(i), it follows that
$\alpha\beta=\sum_{j=0}^re_j(x)(\xi_j\eta_j)$. This implies that
$\alpha\beta=0$ in $R[x]/\langle x^n-1\rangle$ if and only if $\xi_j\eta_j=0$
in $\mathcal{K}_j+v\mathcal{K}_j$ for all $0\leq j\leq r$ by Lemma 3.2(ii).
From this and by $\mathcal{C}=\{\sum_{j=0}^re_j(x)\xi_j\mid \xi_j\in C_j, \ j=0,1,\ldots,r\}$, we deduce that
\begin{eqnarray*}
 {\rm Ann}(\mathcal{C})
  &=&\{\beta\in R[x]/\langle x^n-1\rangle\mid \alpha\beta=0, \forall \alpha\in \mathcal{C}\}\\
  &=&\{\sum_{j=0}^re_j(x)\eta_j\mid \eta_j\in \mathcal{K}_j+v\mathcal{K}_j, \ \xi_j\eta_j=0, \
  \forall  \xi_j\in C_j, \ j=0,1,\ldots,r \}\\
  &=&\{\sum_{j=0}^re_j(x)\eta_j\mid \eta_j\in {\rm Ann}(C_j), \ j=0,1,\ldots,r\}\\
  &=&\bigoplus_{j=0}^re_j(x){\rm Ann}(C_j) \ ({\rm mod} \ x^n-1).
\end{eqnarray*}
Since $\tau$ is a ring isomorphism, by Lemma 4.1 it follows that
\begin{eqnarray*}
{\cal C}^{\bot}&=&\tau({\rm Ann}({\cal C}))=\tau(\sum_{j=0}^re_j(x){\rm Ann}(C_j))\\
 &=&\{\sum_{j=0}^r\tau(e_j(x)a_j(x)+ve_j(x)b_j(x))\mid a_j(x), b_j(x)\in \mathcal{K}_j, \\
 && a_j(x)+vb_j(x)\in {\rm Ann}(C_j), \ j=0,1,\ldots,r\} \\
 &=&\{\sum_{j=0}^r(e_j(x^{-1})a_j(x^{-1})+ve_j(x^{-1})b_j(x^{-1}))\mid a_j(x), b_j(x)\in \mathcal{K}_j, \\
 && a_j(x)+vb_j(x)\in {\rm Ann}(C_j), \ j=0,1,\ldots,r\} \\
 &=&\sum_{j=0}^re_j(x^{-1})\{\sum_{j=0}^r(a_j(x^{-1})+vb_j(x^{-1}))\mid a_j(x), b_j(x)\in \mathcal{K}_j, \\
 && a_j(x)+vb_j(x)\in {\rm Ann}(C_j), \ j=0,1,\ldots,r\} \\
 &=&\bigoplus_{j=0}^re_j(x^{-1})\tau({\rm Ann}(C_j)) \ ({\rm mod} \ x^n-1).
\end{eqnarray*}
Finally, as $x^n-1=0$ in $\mathcal{A}$, we have that $x^{n}=1$ and $x^{-1}=x^{n-1}$ (mod $x^n-1$) in the ring
$\mathcal{A}+v\mathcal{A}$.
\hfill
$\Box$

\vskip 3mm \noindent
  {\bf Remark} For each integer $j$, $0\leq j\leq r$, and
an ideal $C_j$ of $\mathcal{K}_j+v\mathcal{K}_j$, the annihilating ideal
${\rm Ann}(C_j)$ of $C_j$ has been  expressed explicitly by Theorem 3.5. Hence
for each cyclic code over $R$ of length $n$
with a canonical form decomposition ${\cal C}=\bigoplus_{j=0}^re_j(x)C_j$, its
dual code can be completely determined by Theorems 4.2 and 3.5.

\vskip 3mm \par
   Next, we consider how to determine all self-dual cyclic codes over $R$ of length $n$.

\par
  For any polynomial $f(x)=\sum_{i=0}^ma_ix^i\in \mathbb{Z}_{p^s}[x]$
of degree $m\geq 0$. The \textit{reciprocal polynomial} of $f(x)$ is defined
by $\widetilde{f}(x)=x^mf(\frac{1}{x})$, and $f(x)$ is said to be \textit{self-reciprocal}
if $\widetilde{f}(x)=\delta f(x)$ for some $\delta\in \mathbb{Z}_{p^s}^\times$.
By Equation (3) in Section 3, we have $x^n-1=f_0(x)f_1(x)\ldots f_r(x)$. This implies
$$x^n-1=(-1)\widetilde{f}_0(x)\widetilde{f}_1(x)\ldots \widetilde{f}_r(x).$$
 Since $f_0(x),f_1(x),\ldots, f_r(x)$ are pairwise coprime monic basic irreducible polynomials in
$\mathbb{Z}_{p^s}[x]$, $\widetilde{f}_0(x)\widetilde{f}_1(x)\ldots \widetilde{f}_r(x)$
are pairwise coprime basic irreducible polynomials in
$\mathbb{Z}_{p^s}[x]$ as well. Hence for any integer $j$, $0\leq j\leq r$, there
is a unique integer $j^\prime$ such that
$$\widetilde{f}_j(x)=\delta_jf_{j^\prime}(x) \ {\rm where} \ \delta_j\in \mathbb{Z}_{p^s}^\times, \ 0\leq j^\prime\leq r.$$
Especially, we have $0^\prime=0$ by $f_0(x)=x-1$ and $\widetilde{f}_0(x)=1-x=(-1)f_0(x)$.
Then by Equation (5), $x^n=1$ in $\mathcal{A}$ and $x^{m_j}f_j(x^{-1})=\widetilde{f}_j(x)$, it follows that
\begin{eqnarray*}
e_j(x^{-1})&=&1-d_j(x^{-1})f_j(x^{-1})\\
  &=&1-x^{n-{\rm deg}(d_j(x))-m_j}(x^{{\rm deg}(d_j(x))}d_j(x^{-1}))(x^{m_j}f_j(x^{-1}))\\
  &=&1-b_j(x)f_{j^\prime}(x)
\end{eqnarray*}
where $b_j(x)=\delta_jx^{n-{\rm deg}(d_j(x))-m_j}\widetilde{d}_j(x)\in \mathcal{A}$.
Then by $e_j(x)=c_j(x)F_j(x)$ where $F_j(x)=\frac{x^n-1}{f_j(x)}$,
a similar argument shows that $e_j(x^{-1})=a_j(x)F_{j^\prime}(x)$ for some $a_j(x)\in \mathcal{A}$.
Then we have $a_j(x)F_{j^\prime}(x)+b_j(x)f_{j^\prime}(x)=1$. Comparing this with Equation (4), by the definition of $e_{j^\prime}(x)$ we see that $e_j(x^{-1})=e_{j^\prime}(x)$.

\par
  We still use $\tau$ to denote the map $j\mapsto j^\prime$.
Then we have
\begin{equation}
\widetilde{f}_j(x)=\delta_jf_{\tau(j)}(x) \ {\rm and} \ \tau(e_j(x))=e_j(x^{-1})=e_{\tau(j)}(x).
\end{equation}
Whether $\tau$ denotes the ring automorphism of $\mathcal{A}$ or this map  is determined by context. The next lemma shows the compatibility of the two
uses of $\tau$.

\vskip 3mm \noindent
  {\bf Lemma 4.3} \textit{Using the notations above, we have the following conclusions}.

\par
  (i) \textit{$\tau$ is a permutation on on the
set $\{0,1,\ldots,r\}$ satisfying $\tau^{-1}=\tau$}.

\par
  (ii) \textit{After a rearrangement of $f_1(x),\ldots,f_r(x)$, there exists a unique
pair $(\lambda,\rho)$ of nonnegative integers such that
$\lambda+2\rho=r$, $\tau(j)=j$ for all $0\leq j\leq \lambda$ and
$\tau(\lambda+l)=\lambda+l+\rho$ for all $1\leq l\leq \rho$}.

\par
  (iii) \textit{For any integer $j$, $0\leq j\leq r$, the map $\tau_j$ defined by
$$\tau_j(a(x))=a(x^{-1})=a(x^{n-1}) \ ({\rm mod} \ f_{\tau(j)}(x)),  \forall a(x)\in \mathcal{K}_j)$$
is an isomorphism of rings from
$\mathcal{K}_j$ onto $\mathcal{K}_{\tau(j)}$}.

\vskip 3mm \noindent
  {\bf Proof.} (i) It follows from the definition of the map $\tau$ and that
$f_{\tau(\tau(j))}(x)=\delta_{\tau(j)}^{-1}\widetilde{f}_{\tau(j)}(x)=
\delta_{\tau(j)}^{-1}\delta_j^{-1}\widetilde{\widetilde{f}_j}(x)=\delta_{\tau(j)}^{-1}\delta_j^{-1}f_j(x)
=f_j(x)$ by Equation (7).

\par
  (ii) It follows from (i) and the properties of permutations on a finite set.

\par
  (iii) Since $\tau$ is an automorphism of the ring $\mathcal{A}$ and
$\tau(e_j(x))=e_{\tau(j)}(x)$, by Lemma 3.2(ii) we see that the restriction $\tau|_{\mathcal{A}e_j(x)}$
of $\tau$ on $\mathcal{A}e_j(x)$ is an isomorphism of rings from $\mathcal{A}e_j(x)$ onto
$\mathcal{A}e_{\tau(j)}(x)$. By Lemma 3.2(ii), $\varphi_j$ is a ring isomorphism from
$\mathcal{K}_j$ onto $\mathcal{A}e_j(x)$ and $\varphi_{\tau(j)}$ is a ring isomorphism from
$\mathcal{K}_{\tau(j)}$ onto $\mathcal{A}e_{\tau(j)}(x)$. Now, let $\psi=\varphi_{\tau(j)}^{-1}(\tau|_{\mathcal{A}e_j(x)})\varphi_j$. Then $\psi$ must be a ring isomorphism from
$\mathcal{K}_j$ onto $\mathcal{K}_{\tau(j)}$. For any $a(x)\in \mathcal{K}_j$, by Lemma 3.2(ii) and the definition of $\tau|_{\mathcal{A}e_j(x)}$ we have
\begin{eqnarray*}
\psi(a(x))
 &=&\varphi_{\tau(j)}^{-1}(\tau|_{\mathcal{A}e_j(x)})(\varphi_j(a(x)))
 =\varphi_{\tau(j)}^{-1}(\tau|_{\mathcal{A}e_j(x)}(e_j(x)a(x)))\\
 &=&\varphi_{\tau(j)}^{-1}(e_j(x^{-1})a(x^{-1}))=\varphi_{\tau(j)}^{-1}(e_{\tau(j)}(x)a(x^{n-1}))\\
 &=&e_{\tau(j)}(x)a(x^{n-1}) \ ({\rm mod} \ f_{\tau(j)}(x)) \\
 &=&a(x^{n-1}) \ ({\rm mod} \ f_{\tau(j)}(x)) \\
 &=&\tau_j(a(x)),
\end{eqnarray*}
since $e_{\tau(j)}(x)\equiv 1-b_j(x)f_{\tau(j)}(x)\equiv 1$ (mod $f_{\tau(j)}(x)$). This implies
$\tau_j=\psi$.
\hfill
$\Box$

\vskip 3mm \par
  For any integer $0\leq j\leq r$ and an ideal $C_j$ of the ring $\mathcal{K}_j+v\mathcal{K}_j$,
using the notations of Theorem 4.2 and by Lemma 4.3(iii) we see that
$\tau({\rm Ann}(C_j))=\{\tau_j(a(x))+v\tau_j(b(x))\mid a(x),b(x)\in\mathcal{K}_j, \
a(x)+vb(x)\in {\rm Ann}(C_j)\}$. Hence $\tau({\rm Ann}(C_j))$ is an ideal of the ring
$\mathcal{K}_{\tau(j)}+v\mathcal{K}_{\tau(j)}$.

\par
  Now, let $\mathcal{C}$ be a cyclic code over $R$ of length $n$
with a canonical form decomposition ${\cal C}=\bigoplus_{j=0}^{\lambda+2\rho}e_j(x)C_j$, where
$C_j$ is an ideal of $\mathcal{K}_j+v\mathcal{K}_j$. Then by Theorem 4.2, Equation (7) and Lemma 4.3(ii), the dual code of $\mathcal{C}$ is given by
\begin{eqnarray*}
{\cal C}^{\bot}&=&\left(\bigoplus_{0\leq j\leq \lambda} e_{j}(x)\tau({\rm Ann}(C_j))\right)\\
 &&\oplus
\left(\bigoplus_{1\leq l\leq \rho}\left(e_{\lambda+\rho+l}(x)\tau({\rm Ann}(C_{\lambda+l}))\right)
\oplus\left(e_{\lambda+l}(x)\tau({\rm Ann}(C_{\lambda+\rho+l}))\right)\right)
\end{eqnarray*}
$({\rm mod} \ x^n-1)$. From this and by Lemma 4.3(ii), we deduce that
$\mathcal{C}={\cal C}^{\bot}$ if and only if the following two conditions are satisfied:

\vskip 2mm\noindent
  $\bullet$ $C_j=\tau({\rm Ann}(C_j))$, for all $0\leq j\leq \lambda$;

\vskip 2mm\noindent
  $\bullet$ $C_{\lambda+\rho+l}=\tau({\rm Ann}(C_{\lambda+l}))$ which implies
$C_{\lambda+l}=\tau({\rm Ann}(C_{\lambda+\rho+l}))$, where
$C_{\lambda+l}$ is an arbitrary ideal of $\mathcal{K}_{\lambda+l}+v\mathcal{K}_{\lambda+l}$,
for all $1\leq l\leq \rho$.

\vskip 2mm\par
  Finally, by Theorems 3.3 and 3.5 we give all distinct self-dual cyclic codes over $R$ of length $n$ by the following theorem.

\vskip 3mm \noindent
  {\bf Theorem 4.4} \textit{For any integer $0\leq j\leq \lambda$, set}
$$\mathcal{W}_{j}=\{w(x)\in T_{f_j}=\mathbb{Z}_p[x]/\langle \overline{f}_j(x)\rangle\mid w(x)+w(x^{n-1})+1\equiv 0 \
({\rm mod} \ \overline{f}_j(x))\}$$
\textit{and denote $\omega_{j}=|\mathcal{W}_j|$. Let $2\leq s\leq 4$. Then all distinct self-dual cyclic codes over $R=\mathbb{Z}_{p^s}+v\mathbb{Z}_{p^s}$ of length $n$ are given by}:
$${\cal C}=\bigoplus_{0\leq j\leq\lambda+2\rho}e_j(x)C_j \ ({\rm mod} \ x^n-1),$$
\textit{where $C_j$ is an ideal of $\mathcal{K}_j+v\mathcal{K}_j$ given by one the following cases}.

\vskip 2mm\par
  $\bullet$ \textit{Let $s=2$. For each $0\leq j\leq \lambda$, $C_j$ is one of the following $\omega_j+1$ ideals}:

\par
  \textit{$\langle p\rangle$, $\langle pw(x)+v\rangle$ where $w(x)\in \mathcal{W}_{j}$}.

\vskip 2mm\par
  \textit{For each $1\leq l\leq \rho$, the pair $(C_{\lambda+l}, C_{\lambda+\rho+l})$ is given
by the following table}:
\begin{center}
\begin{tabular}{lll|l}\hline
 N     &     $C_{\lambda+l}$ (mod $f_{\lambda+l}(x)$) & $|C_{\lambda+l}|$  & $C_{\lambda+\rho+l}$ (mod $f_{\lambda+\rho+l}(x)$) \\ \hline
$p^{m_{\lambda+l}}$ &  $\langle p w(x)+v\rangle$ & $p^{2m_{\lambda+l}}$  & $\langle p(-1-w(x^{n-1}))+v\rangle$ \\
$1$ &  $\langle pv\rangle$ & $p^{m_{\lambda+l}}$  & $\langle p,v\rangle$ \\
$3$ &  $\langle p^k\rangle$, $0\leq k\leq 2$ & $p^{(4-2k)m_{\lambda+l}}$  & $\langle p^{2-k}\rangle$ \\
$1$ &  $\langle p,v\rangle$ & $p^{3m_{\lambda+l}}$  & $\langle pv\rangle$ \\
\hline
\end{tabular}
\end{center}
\textit{where $w(x)\in T_{f_{\lambda+l}}$ and $N$ is the number of ideals in the same row}.

\par
\textit{Therefore, the number of self-dual cyclic codes over $\mathbb{Z}_{p^2}+v\mathbb{Z}_{p^2}$ $(v^2=pv)$ of length $n$ is equal to $\prod_{0\leq j\leq \lambda}(\omega_j+1)\prod_{1\leq l\leq \rho}(p^{m_{\lambda+l}}+5)$}.

\vskip 2mm\par
  $\bullet$ \textit{Let $s=3$. For each $0\leq j\leq \lambda$, $C_j=\langle p^2,pv\rangle$}.

\vskip 2mm\par
  \textit{For each $1\leq l\leq \rho$, the pair $(C_{\lambda+l}, C_{\lambda+\rho+l})$ is given
by the following table}:
\begin{center}
\begin{tabular}{lll|l}\hline
 N     &     $C_{\lambda+l}$  (mod $f_{\lambda+l}(x)$) & $|C_j|$  & $C_{\lambda+\rho+l}$ (mod $f_{\lambda+\rho+l}(x)$)  \\ \hline
$p^{m_{\lambda+l}}$ &  $\langle p^2 w(x)+v\rangle$ & $p^{3m_{\lambda+l}}$  & $\langle -p-p^2w(x^{n-1}) +v\rangle$ \\
$p^{m_{\lambda+l}}$ &  $\langle -p+p^2 w(x)+v\rangle$ & $p^{3m_{\lambda+l}}$  & $\langle -p^2w(x^{n-1})+v\rangle$ \\
$1$ &  $\langle p^2v\rangle$ & $p^{m_{\lambda+l}}$  & $\langle p,v\rangle$ \\
$p^{m_{\lambda+l}}$ &  $\langle p^2w(x)+pv\rangle$ & $p^{2m_{\lambda+l}}$  & $\langle p^2, p(-1-w(x^{n-1}))+v\rangle$ \\
$4$ &  $\langle p^k\rangle$, $0\leq k\leq 3$ & $p^{(6-2k)m_{\lambda+l}}$  & $\langle p^{3-k}\rangle$ \\
$1$ &  $\langle p,v\rangle$ & $p^{5m_{\lambda+l}}$  & $\langle p^2v\rangle$ \\
$p^{m_{\lambda+l}}$ &  $\langle p^2,pw(x)+v\rangle$ & $p^{4m_{\lambda+l}}$  & $\langle p^2(-1-w(x^{n-1}))+pv\rangle$ \\
$1$ &  $\langle p^2,pv\rangle$ & $p^{3m_{\lambda+l}}$  & $\langle p^2,pv\rangle$ \\
\hline
\end{tabular}
\end{center}
\textit{where $w(x)\in T_{f_{\lambda+l}}$ and $N$ is the number of ideals in the same row}.

\par
\textit{Therefore, the number of self-dual cyclic codes over $\mathbb{Z}_{p^3}+v\mathbb{Z}_{p^3}$ $(v^2=pv)$ of length $n$ is equal to $\prod_{1\leq l\leq \rho}(4p^{m_{\lambda+l}}+7)$}.

\vskip 2mm\par
  $\bullet$ \textit{Let $s=4$. For each $0\leq j\leq \lambda$, $C_j$ is one of the following $\omega_j+1$ ideals}:

\par
  \textit{$\langle p^2\rangle$, $\langle p^3, p^2w(x)+pv\rangle$ where $w(x)\in \mathcal{W}_{j}$}.

\vskip 2mm\par
  \textit{For each $1\leq l\leq \rho$, the pair $(C_{\lambda+l}, C_{\lambda+\rho+l})$ is given
by the following table}:
{\small \begin{center}
\begin{tabular}{lll|l}\hline
 N     &     $C_{\lambda+l}$ (mod $f_{\lambda+l}(x)$) & $|C_{\lambda+l}|$  & $C_{\lambda+\rho+l}$ (mod $f_{\lambda+\rho+l}(x)$)   \\ \hline
$p^{m_{\lambda+l}}$ &  $\langle p^3 w(x)+v\rangle$ & $p^{4m_{\lambda+l}}$  & $\langle -p-p^3w(x^{n-1})+v\rangle$ \\
$p^{m_{\lambda+l}}$ &  $\langle -p+p^3 w(x)+v\rangle$ & $p^{4m_{\lambda+l}}$  & $\langle -p^3w(x^{n-1})+v\rangle$ \\
$1$ &  $\langle p^3v\rangle$ & $p^{m_{\lambda+l}}$  & $\langle p,v\rangle$ \\
$p^{m_{\lambda+l}}$ &  $\langle p^3w(x)+p^2v\rangle$ & $p^{2m_{\lambda+l}}$  & $\langle p^2, p(-1-w(x^{n-1}))+v\rangle$ \\
$p^{m_{\lambda+l}}$ &  $\langle p^3w(x)+pv\rangle$ & $p^{3m_{\lambda+l}}$  & $\langle p^3, -p-p^2w(x^{n-1})+v\rangle$ \\
$p^{m_{\lambda+l}}$ &  $\langle -p^2+p^3w(x)+pv\rangle$ & $p^{3m_{\lambda+l}}$  & $\langle p^3, -p^2w(x^{n-1})+v\rangle$ \\
$5$ &  $\langle p^k\rangle$, $0\leq k\leq 4$ & $p^{(8-2k)m_{\lambda+l}}$  & $\langle p^{4-k}\rangle$ \\
$1$ &  $\langle p,v\rangle$ & $p^{7m_{\lambda+l}}$  & $\langle p^3v\rangle$ \\
$p^{m_{\lambda+l}}$ &  $\langle p^2,pw(x)+v\rangle$ & $p^{6m_{\lambda+l}}$  & $\langle p^3(-1-w(x^{n-1}))+p^2v\rangle$ \\
$p^{m_{\lambda+l}}$ &  $\langle p^3,p^2w(x)+v\rangle$ & $p^{5m_{\lambda+l}}$  & $\langle -p^2-p^3w(x^{n-1})+pv\rangle$ \\
$p^{m_{\lambda+l}}$ &  $\langle p^3,-p+p^2w(x)+v\rangle$ & $p^{5m_{\lambda+l}}$  & $\langle -p^3w(x^{n-1})+pv\rangle$ \\
$1$ &  $\langle p^2,pv\rangle$ & $p^{5m_{\lambda+l}}$  & $\langle p^3,p^2v\rangle$ \\
$1$ &  $\langle p^3,p^2v\rangle$ & $p^{3m_{\lambda+l}}$  & $\langle p^2,pv\rangle$ \\
$p^{m_{\lambda+l}}$ &  $\langle p^3,p^2w(x)+pv\rangle$ & $p^{4m_{\lambda+l}}$  & $\langle p^3,p^2(-1-w(x^{n-1}))+pv\rangle$ \\
\hline
\end{tabular}
\end{center} }
\noindent
\textit{where $w(x)\in T_{f_{\lambda+l}}$ and $N$ is the number of ideals in the same row}.

\par
\textit{Therefore, the number of self-dual cyclic codes over $\mathbb{Z}_{p^4}+v\mathbb{Z}_{p^4}$ $(v^2=pv)$ of length $n$ is equal to $\prod_{0\leq j\leq \lambda}(\omega_j+1)\prod_{1\leq l\leq \rho}(9p^{m_{\lambda+l}}+9)$}.

\vskip 3mm \noindent
  {\bf Remark} Let $s\geq 5$. By Theorem 3.5, one can easily list all distinct self-dual cyclic codes over $R=\mathbb{Z}_{p^s}+v\mathbb{Z}_{p^s}$ $(v^2=pv)$  of length $n$. Let $\Lambda(p,s,n)$ be the number of self-dual cyclic codes over $R$ of length $n$. Then

\par
  $\Lambda(p,s,n)=\prod_{1\leq l\leq \rho}((s-1)^2p^{m_{\lambda+l}}+2s+1)$ when $s$ is odd,
and

\par
  $\Lambda(p,s,n)=\prod_{0\leq j\leq \lambda}(\omega_j+1)\prod_{1\leq l\leq \rho}((s-1)^2p^{m_{\lambda+l}}+2s+1)$ when $s$ is even.



\section{Cyclic codes over $\mathbb{Z}_{4}+v\mathbb{Z}_{4}$}
\noindent
  In this section, we consider cyclic codes over $R=\mathbb{Z}_4[v]/\langle v^2+2v\rangle=\mathbb{Z}_4+v\mathbb{Z}_4$ $(v^2=2v)$.
  As in [13] Section 3, we define $\varrho: R\rightarrow \mathbb{Z}_4^2$ by
$$\varrho(\alpha)=(a+b,b), \ \forall \alpha=a+bv\in R \ {\rm where} \ a,b\in \mathbb{Z}_4$$
and let $\theta: R^n\rightarrow \mathbb{Z}_4^{2n}$ be such that
$$\theta(\alpha_1,\ldots,\alpha_n)=(\varrho(\alpha_1),\ldots,\varrho(\alpha_n)),
\ \forall \alpha_1,\ldots,\alpha_n\in R.$$
Let $w_L$ denote the Lee weight on $\mathbb{Z}_4$ defined by: $w_L(0)=0$, $w_L(1)=w_L(3)=1$
and $w_L(2)=2$. We extend $w_L$ in a natural way. For $a+bv\in R$ with $a,b\in \mathbb{Z}_4$, define
$$w_L(a+bv)=w_L(a+b)+w_L(b).$$
With this distance and Gray map definition, the following conclusions have been verified by Mart\'{\i}nez-Moro et al. [13].

\vskip 3mm \noindent
  {\bf Lemma 5.1} ([13] Theorem 3.1) \textit{Let $\mathcal{C}$ be a linear code over $R$ of length $n$ and minimum Lee distance $d$. Then $\theta(\mathcal{C})$ is a linear code over $\mathbb{Z}_4$ of length $2n$, $|\theta(\mathcal{C})|=|\mathcal{C}|$ and is of minimum Lee distance $d$}.

\vskip 3mm \noindent
  {\bf Lemma 5.2} ([13] Proposition 3.3) \textit{Let $\mathcal{C}$ be a linear
code over $R$ of length $n$. Then} $\theta(\mathcal{C}^{\bot})=\theta(\mathcal{C})^{\bot}$.
\textit{In particular, if $\mathcal{C}$ is self-dual, then $\theta(\mathcal{C})$ is an
self-dual code over $\mathbb{Z}_4$ of length $2n$ and has
the same Lee weight distribution}.

\vskip 3mm \par
   Moreover, we have the following properties for cyclic codes $R$ of length $n$.

\vskip 3mm \noindent
  {\bf Proposition 5.3} \textit{Let $\mathcal{C}$ be a cyclic codes $R$ of length $n$.
Then $\theta(\mathcal{C})$ is a $2$-quasi-cyclic code over $\mathbb{Z}_4$ of length $2n$}.

\vskip 3mm \noindent
  {\bf Proof.} Let $\underline{\alpha}=(\alpha_0,\alpha_1,\ldots,\alpha_{n-1})\in \mathcal{C}$,
where $\alpha_i=a_i+b_iv$ with $a_i,b_i\in \mathbb{Z}_4$ for all $i=0,1,\ldots,n-1$. Then
$\theta(\underline{\alpha})=(a_0+b_0,b_0,a_1+b_1,b_1,\ldots,a_{n-1}+b_{n-1}, b_{n-1})\in \theta(\mathcal{C})$. Since $\mathcal{C}$ is cyclic, we have $(\alpha_{n-1},\alpha_0,\alpha_1,\ldots,\alpha_{n-2})\in \mathcal{C}$. This implies
$(a_{n-1}+b_{n-1}, b_{n-1},a_0+b_0,b_0,a_1+b_1,b_1,\ldots,a_{n-2}+b_{n-2}, b_{n-2})\in \theta(\mathcal{C})$. Hence $\theta(\mathcal{C})$ is a $2$-quasi-cyclic code over $\mathbb{Z}_4$ of length $2n$.
\hfill
$\Box$

\vskip 3mm
\par
   As an example, we consider cyclic codes over $R$ of length $15$. In this case, $p=s=2$ and
$x^{15}-1=f_0(x)f_1(x)f_2(x)f_3(x)f_4(x)$ where $f_0(x)=x-1$,
$f_1(x)=x^2+x+1$, $f_2(x)=x^4+x^3+x^2+x+1$, $f_3(x)=x^4+2x^2+3x+1$ and $f_4(x)=\widetilde{f}_3(x)$ are
monic basic irreducible polynomials in $\mathbb{Z}_4[x]$. Hence $m_1=2$ and
$m_2=m_3=m_4=4$. By Corollary 3.6, the number of cyclic codes over $\mathbb{Z}_4+v\mathbb{Z}_4$
of length $15$ is equal to
$$(2^1+5)\cdot(2^2+5)\cdot(2^4+5)^3=583443.$$
Now, for each integer $j$, $0\leq j\leq 4$, we denote $F_j(x)=\frac{x^{15}-1}{f_j(x)}$,
$$\mathcal{K}_j=\mathbb{Z}_4[x]/\langle f_j(x)\rangle=\{\sum_{i=0}^{m_i-1}a_ix^i\mid a_0,a_1,\ldots,a_{m_i-1}\in \mathbb{Z}_4\}$$
and $T_{f_j}=\{\sum_{i=0}^{m_i-1}b_ix^i\mid b_0,b_1,\ldots,b_{m_i-1}\in \mathbb{Z}_2\}$.
Then we find polynomials
$c_j(x),d_j(x)\in \mathbb{Z}_4[x]$ satisfying $c_j(x)F_j(x)+d_j(x)f_j(x)=1$ and set
$e_j(x)\in \mathbb{Z}_4[x]/\langle x^{15}-1\rangle$ such that $e_j(x)\equiv c_j(x)F_j(x)$ (mod $x^{15}-1$).
Precisely, we have

\par
 $e_0(x)=3\,{x}^{14}+3\,{x}^{13}+3\,{x}^{12}+3\,{x}^{11}+3\,{x}^{10}+3\,{x}^{9}
+3\,{x}^{8}+3\,{x}^{7}+3\,{x}^{6}+3\,{x}^{5}+3\,{x}^{4}+3\,{x}^{3}+3\,
{x}^{2}+3\,x+3$,

\par
  $e_1(x)={x}^{14}+{x}^{13}+2\,{x}^{12}+{x}^{11}+{x}^{10}+2\,{x}^{9}+{x}^{8}+{x}
^{7}+2\,{x}^{6}+{x}^{5}+{x}^{4}+2\,{x}^{3}+{x}^{2}+x+2$,

\par
  $e_2(x)={x}^{14}+{x}^{13}+{x}^{12}+{x}^{11}+{x}^{9}+{x}^{8}+{x}^{7}+{x}^{6}+{x
}^{4}+{x}^{3}+{x}^{2}+x$,

\par
  $e_3(x)={x}^{12}+2\,{x}^{10}+{x}^{9}+3\,{x}^{8}+{x}^{6}+2\,{x}^{5}+3\,{x}^{4}+
{x}^{3}+3\,{x}^{2}+3\,x$,

\par
  $e_4(x)=3\,{x}^{14}+3\,{x}^{13}+{x}^{12}+3\,{x}^{11}+2\,{x}^{10}+{x}^{9}+3\,{x
}^{7}+{x}^{6}+2\,{x}^{5}+{x}^{3}$.

\vskip 3mm\par
  $\diamondsuit$ By Theorem 3.3(ii) and Theorem 3.5, all distinct $583443$
cyclic codes over $R$
of length $15$ are given by
$$\mathcal{C}=e_0(x)C_0\oplus e_1(x)C_1\oplus e_2(x)C_2\oplus e_3(x)C_3\oplus e_4(x)C_4,$$
where

\noindent
  $\bullet$ $C_0$ is one of the following $7$ ideals of $\mathbb{Z}_4+v\mathbb{Z}_4$:

\par
   $C_0=\langle 2a+v\rangle$ with $|C_0|=4$, where $a\in \mathbb{Z}_2=\{0,1\}$;
\par
 $C_0=\langle 2^k\rangle$  with $|C_0|=2^{4-2k}$, where $0\leq k\leq 2$;

\par
 $C_0=\langle 2v\rangle$  with $|C_0|=2$; $C_0=\langle 2,v\rangle$ with $|C_0|=8$.

\noindent
  $\bullet$ $C_1$ is one of the following $9$ ideals of $\mathcal{K}_1+v\mathcal{K}_1$:

\par
   $C_1=\langle 2(a_0+a_1x)+v\rangle$ with $|C_1|=16$, where $a_0,a_1\in \mathbb{Z}_2=\{0,1\}$;
\par
 $C_1=\langle 2^k\rangle$  with $|C_1|=4^{4-2k}$, where $0\leq k\leq 2$;

\par
 $C_1=\langle 2v\rangle$  with $|C_1|=4$; $C_1=\langle 2,v\rangle$ with $|C_1|=64$.

\noindent
  $\bullet$ $C_j$ is one of the following $21$ ideals of $\mathcal{K}_j+v\mathcal{K}_j$ for all $j=2,3,4$:

\par
   $C_j=\langle 2(a_0+a_1x+a_2x^2+a_3x^3)+v\rangle$ with $|C_j|=16^2$, where $a_0,a_1,a_2,a_3\in \mathbb{Z}_2=\{0,1\}$;
 $C_j=\langle 2^k\rangle$  with $|C_j|=16^{4-2k}$, where $0\leq k\leq 2$;

\par
 $C_j=\langle 2v\rangle$  with $|C_j|=16$; $C_j=\langle 2,v\rangle$ with $|C_j|=16^3$.

\vskip 3mm\par
  $\diamondsuit$  Using the notations of Section 4, we have $r=4$, $\lambda=2$ and $\rho=1$.
By $T_{f_j}=\mathbb{Z}_2[x]/\langle \overline{f}_j(x)\rangle=\{\sum_{i=0}^{m_i-1}a_ix^i\mid a_0,a_1,\ldots,a_{m_i-1}\in \mathbb{Z}_2\}$, we have the following

\par
  $\mathcal{W}_0=\{a\in \mathbb{Z}_2\mid a+a+1\equiv 0 \ ({\rm mod} \ \overline{f}_0(x))\}=\emptyset$. This implies
$\omega_0=0$.

\par
  $\mathcal{W}_1=\{a+bx\in \mathbb{Z}_2[x]/\langle \overline{f}_1(x)\rangle\mid (a+bx)+(a+bx^{14})+1\equiv 0 \ ({\rm mod} \ \overline{f}_1(x))\}=\{a+bx\mid a+bx+a+b(1+x)+1=0, \ a,b\in \mathbb{Z}_2\}=\{x,1+x\}$. This implies
$\omega_1=2$.

\par
  $\mathcal{W}_2=\{a+bx+cx^2+dx^3\in \mathbb{Z}_2[x]/\langle \overline{f}_2(x)\rangle\mid (a+bx+cx^2+dx^3)+(a+bx^{14}+cx^{13}+dx^{12})+1\equiv 0 \ ({\rm mod} \ \overline{f}_2(x))\}=\{a+bx+cx^2+dx^3\mid bx+cx^2+dx^3+b(1+x+x^2+x^3)+cx^3+dx^2+1=0, \ a,b,c,d\in \mathbb{Z}_2\}=\{a+x+cx^2+(1+c)x^3\mid
a,c\in \mathbb{Z}_2\}$. This implies
$\omega_2=4$.

\par
  By Theorem 4.4, the number of all distinct self-dual cyclic codes over $R$ of length $15$ is equal to
$$(0+1)\cdot(2+1)\cdot(4+1)\cdot(2^4+5)=315.$$
Precisely, all these $315$ codes are given by: $\mathcal{C}=\bigoplus_{0\leq j\leq 4}e_j(x)C_j$, where

\par
   $C_0=\langle 2\rangle=2(\mathbb{Z}_4+v\mathbb{Z}_4)$.

\par
   $C_1=\langle 2\rangle$, $C_1=\langle 2w(x)+v\rangle$ where $w(x)\in\mathcal{W}_1$.

\par
   $C_2=\langle 2\rangle$, $C_2=\langle 2w(x)+v\rangle$ where $w(x)\in\mathcal{W}_2$.

\par
  $(C_3,C_4)$ is given by the following table:
\begin{center}
\begin{tabular}{lll|l}\hline
 N     &     $C_3$ (mod $f_3(x)$) & $|C_4|$  & $C_4$ (mod $f_4(x)$) \\ \hline
$16$ &  $\langle 2 w(x)+v\rangle$ & $2^{8}$  & $\langle 2(1+w(x^{14}))+v\rangle$ \\
$1$ &  $\langle 2v\rangle$ & $2^{4}$  & $\langle 2,v\rangle$ \\
$3$ &  $\langle 2^k\rangle$, $0\leq k\leq 2$ & $2^{4(4-2k)}$  & $\langle 2^{2-k}\rangle$ \\
$1$ &  $\langle 2,v\rangle$ & $2^{12}$  & $\langle 2v\rangle$ \\
\hline
\end{tabular}
\end{center}
in which $w(x)=a+bx+cx^2+dx^3$ and
$w(x^{14})=a+d+(d+c)x+(c+b)x^2+bx^3$ (mod $f_4(x)$, mod $2$) where $a,b,c,d\in \mathbb{Z}_2$, and $N$ is the number of ideals in the same row.

\vskip 3mm
  \par
By Proposition 5.3, we obtain $315$ $2$-quasi-cyclic self-dual codes over $\mathbb{Z}_4$ of length $30$. Among these codes, there are $70$ codes with minimum Lee weight $12$.
These $70$ $2$-quasi-cyclic self-dual codes over $\mathbb{Z}_4$ are given by the following table, with $C_1$, $C_2$, $C_3$, $C_4$ and the type of each $\mathbb{Z}_4$-code $\theta(\mathcal{C})$.

\begin{center}\tiny
\begin{tabular}{llll|l}\hline
  $C_1$     & $C_2$     &     $C_3$ (mod $f_3(x)$) & $C_4$ (mod $f_4(x)$)& Type\\ \hline
$\langle     v + 2x\rangle $ & $\langle     2x^3 + 2x + v\rangle $ & $\langle                           2\rangle $ & $\langle                           2\rangle $ & $ 2^{18} 4^6$\\
$\langle v + 2x + 2\rangle $ & $\langle     2x^3 + 2x + v\rangle $ & $\langle                           2\rangle $ & $\langle                           2\rangle $ & $ 2^{18} 4^6$\\
$\langle     v + 2x\rangle $ & $\langle     2x^2 + 2x + v\rangle $ & $\langle                           2\rangle $ & $\langle                           2\rangle $ & $ 2^{18} 4^6$\\
$\langle v + 2x + 2\rangle $ & $\langle     2x^2 + 2x + v\rangle $ & $\langle                           2\rangle $ & $\langle                           2\rangle $ & $ 2^{18} 4^6$\\
$\langle     v + 2x\rangle $ & $\langle 2x^3 + 2x + v + 2\rangle $ & $\langle                           2\rangle $ & $\langle                           2\rangle $ & $ 2^{18} 4^6$\\
$\langle v + 2x + 2\rangle $ & $\langle 2x^3 + 2x + v + 2\rangle $ & $\langle                           2\rangle $ & $\langle                           2\rangle $ & $ 2^{18} 4^6$\\
$\langle     v + 2x\rangle $ & $\langle 2x^2 + 2x + v + 2\rangle $ & $\langle                           2\rangle $ & $\langle                           2\rangle $ & $ 2^{18} 4^6$\\
$\langle v + 2x + 2\rangle $ & $\langle 2x^2 + 2x + v + 2\rangle $ & $\langle                           2\rangle $ & $\langle                           2\rangle $ & $ 2^{18} 4^6$\\
$\langle     v + 2x\rangle $ & $\langle                   2\rangle $ & $\langle                         2v\rangle $ & $\langle                       2,v\rangle $ & $ 2^{18} 4^6$\\
$\langle v + 2x + 2\rangle $ & $\langle                   2\rangle $ & $\langle                         2v\rangle $ & $\langle                       2,v\rangle $ & $ 2^{18} 4^6$\\
$\langle     v + 2x\rangle $ & $\langle                   2\rangle $ & $\langle                       2,v\rangle $ & $\langle                         2v\rangle $ & $ 2^{18} 4^6$\\
$\langle v + 2x + 2\rangle $ & $\langle                   2\rangle $ & $\langle                       2,v\rangle $ & $\langle                         2v\rangle $ & $ 2^{18} 4^6$\\
$\langle     v + 2x\rangle $ & $\langle     2x^3 + 2x + v\rangle $ & $\langle                       2,v\rangle $ & $\langle                         2v\rangle $ & $ 2^{10} 4^{10}$\\
$\langle     v + 2x\rangle $ & $\langle     2x^2 + 2x + v\rangle $ & $\langle                         2v\rangle $ & $\langle                       2,v\rangle $ & $ 2^{10} 4^{10}$\\
$\langle     v + 2x\rangle $ & $\langle 2x^3 + 2x + v + 2\rangle $ & $\langle                       2,v\rangle $ & $\langle                         2v\rangle $ & $ 2^{10} 4^{10}$\\
$\langle     v + 2x\rangle $ & $\langle 2x^2 + 2x + v + 2\rangle $ & $\langle                         2v\rangle $ & $\langle                       2,v\rangle $ & $ 2^{10} 4^{10}$\\
$\langle     v + 2x\rangle $ & $\langle                   2\rangle $ & $\langle           2x^3 + 2x^2 + v\rangle $ & $\langle                   2x^2 + v\rangle $ & $ 2^{10} 4^{10}$\\
$\langle     v + 2x\rangle $ & $\langle                   2\rangle $ & $\langle             2x^3 + 2x + v\rangle $ & $\langle     2x^3 + 2x^2 + 2x + v\rangle $ & $ 2^{10} 4^{10}$\\
$\langle     v + 2x\rangle $ & $\langle                   2\rangle $ & $\langle     2x^3 + 2x^2 + 2x + v\rangle $ & $\langle                   2x^3 + v\rangle $ & $ 2^{10} 4^{10}$\\
$\langle     v + 2x\rangle $ & $\langle                   2\rangle $ & $\langle               2x^3 + v + 2\rangle $ & $\langle                 v + 2x + 2\rangle $ & $ 2^{10} 4^{10}$\\
$\langle     v + 2x\rangle $ & $\langle                   2\rangle $ & $\langle                           1\rangle $ & $\langle                           0\rangle $ & $ 2^{10} 4^{10}$\\
$\langle v + 2x + 2\rangle $ & $\langle                   2\rangle $ & $\langle                           1\rangle $ & $\langle                           0\rangle $ & $ 2^{10} 4^{10}$\\
$\langle     v + 2x\rangle $ & $\langle                   2\rangle $ & $\langle                           0\rangle $ & $\langle                           1\rangle $ & $ 2^{10} 4^{10}$\\
$\langle v + 2x + 2\rangle $ & $\langle                   2\rangle $ & $\langle                           0\rangle $ & $\langle                           1\rangle $ & $ 2^{10} 4^{10}$\\
$\langle           2\rangle $ & $\langle     2x^3 + 2x + v\rangle $ & $\langle               2x^3 + v + 2\rangle $ & $\langle                 v + 2x + 2\rangle $ & $ 2^6 4^{12}$\\
$\langle           2\rangle $ & $\langle     2x^3 + 2x + v\rangle $ & $\langle 2x^3 + 2x^2 + 2x + v + 2\rangle $ & $\langle               2x^3 + v + 2\rangle $ & $ 2^6 4^{12}$\\
$\langle           2\rangle $ & $\langle     2x^3 + 2x + v\rangle $ & $\langle                           1\rangle $ & $\langle                           0\rangle $ & $ 2^6 4^{12}$\\
$\langle           2\rangle $ & $\langle     2x^2 + 2x + v\rangle $ & $\langle             2x^3 + 2x + v\rangle $ & $\langle     2x^3 + 2x^2 + 2x + v\rangle $ & $ 2^6 4^{12}$\\
$\langle           2\rangle $ & $\langle     2x^2 + 2x + v\rangle $ & $\langle       2x^3 + 2x^2 + v + 2\rangle $ & $\langle               2x^2 + v + 2\rangle $ & $ 2^6 4^{12}$\\
$\langle           2\rangle $ & $\langle     2x^2 + 2x + v\rangle $ & $\langle                           1\rangle $ & $\langle                           0\rangle $ & $ 2^6 4^{12}$\\
$\langle           2\rangle $ & $\langle     2x^2 + 2x + v\rangle $ & $\langle                           0\rangle $ & $\langle                           1\rangle $ & $ 2^6 4^{12}$\\
$\langle           2\rangle $ & $\langle 2x^3 + 2x + v + 2\rangle $ & $\langle                   2x^3 + v\rangle $ & $\langle                     v + 2x\rangle $ & $ 2^6 4^{12}$\\
$\langle           2\rangle $ & $\langle 2x^3 + 2x + v + 2\rangle $ & $\langle     2x^3 + 2x^2 + 2x + v\rangle $ & $\langle                   2x^3 + v\rangle $ & $ 2^6 4^{12}$\\
$\langle           2\rangle $ & $\langle 2x^3 + 2x + v + 2\rangle $ & $\langle                           1\rangle $ & $\langle                           0\rangle $ & $ 2^6 4^{12}$\\
$\langle           2\rangle $ & $\langle 2x^2 + 2x + v + 2\rangle $ & $\langle           2x^3 + 2x^2 + v\rangle $ & $\langle                   2x^2 + v\rangle $ & $ 2^6 4^{12}$\\
$\langle           2\rangle $ & $\langle 2x^2 + 2x + v + 2\rangle $ & $\langle         2x^3 + 2x + v + 2\rangle $ & $\langle 2x^3 + 2x^2 + 2x + v + 2\rangle $ & $ 2^6 4^{12}$\\
$\langle           2\rangle $ & $\langle 2x^2 + 2x + v + 2\rangle $ & $\langle                           1\rangle $ & $\langle                           0\rangle $ & $ 2^6 4^{12}$\\
$\langle           2\rangle $ & $\langle 2x^2 + 2x + v + 2\rangle $ & $\langle                           0\rangle $ & $\langle                           1\rangle $ & $ 2^6 4^{12}$\\
$\langle v + 2x + 2\rangle $ & $\langle     2x^3 + 2x + v\rangle $ & $\langle                           v\rangle $ & $\langle                       v + 2\rangle $ & $ 2^2 4^{14}$\\
$\langle     v + 2x\rangle $ & $\langle     2x^3 + 2x + v\rangle $ & $\langle           2x^3 + 2x^2 + v\rangle $ & $\langle                   2x^2 + v\rangle $ & $ 2^2 4^{14}$\\
$\langle v + 2x + 2\rangle $ & $\langle     2x^3 + 2x + v\rangle $ & $\langle           2x^3 + 2x^2 + v\rangle $ & $\langle                   2x^2 + v\rangle $ & $ 2^2 4^{14}$\\
$\langle     v + 2x\rangle $ & $\langle     2x^3 + 2x + v\rangle $ & $\langle                       v + 2\rangle $ & $\langle                           v\rangle $ & $ 2^2 4^{14}$\\
$\langle v + 2x + 2\rangle $ & $\langle     2x^3 + 2x + v\rangle $ & $\langle               2x^2 + v + 2\rangle $ & $\langle             2x^2 + 2x + v\rangle $ & $ 2^2 4^{14}$\\
$\langle     v + 2x\rangle $ & $\langle     2x^3 + 2x + v\rangle $ & $\langle 2x^3 + 2x^2 + 2x + v + 2\rangle $ & $\langle               2x^3 + v + 2\rangle $ & $ 2^2 4^{14}$\\
$\langle     v + 2x\rangle $ & $\langle     2x^3 + 2x + v\rangle $ & $\langle                           1\rangle $ & $\langle                           0\rangle $ & $ 2^2 4^{14}$\\
$\langle v + 2x + 2\rangle $ & $\langle     2x^3 + 2x + v\rangle $ & $\langle                           1\rangle $ & $\langle                           0\rangle $ & $ 2^2 4^{14}$\\
$\langle     v + 2x\rangle $ & $\langle     2x^2 + 2x + v\rangle $ & $\langle                           v\rangle $ & $\langle                       v + 2\rangle $ & $ 2^2 4^{14}$\\
$\langle     v + 2x\rangle $ & $\langle     2x^2 + 2x + v\rangle $ & $\langle                     v + 2x\rangle $ & $\langle       2x^3 + 2x^2 + v + 2\rangle $ & $ 2^2 4^{14}$\\
$\langle v + 2x + 2\rangle $ & $\langle     2x^2 + 2x + v\rangle $ & $\langle                     v + 2x\rangle $ & $\langle       2x^3 + 2x^2 + v + 2\rangle $ & $ 2^2 4^{14}$\\
$\langle     v + 2x\rangle $ & $\langle     2x^2 + 2x + v\rangle $ & $\langle             2x^3 + 2x + v\rangle $ & $\langle     2x^3 + 2x^2 + 2x + v\rangle $ & $ 2^2 4^{14}$\\
$\langle v + 2x + 2\rangle $ & $\langle     2x^2 + 2x + v\rangle $ & $\langle                       v + 2\rangle $ & $\langle                           v\rangle $ & $ 2^2 4^{14}$\\
$\langle     v + 2x\rangle $ & $\langle     2x^2 + 2x + v\rangle $ & $\langle               2x^3 + v + 2\rangle $ & $\langle                 v + 2x + 2\rangle $ & $ 2^2 4^{14}$\\
$\langle v + 2x + 2\rangle $ & $\langle     2x^2 + 2x + v\rangle $ & $\langle               2x^3 + v + 2\rangle $ & $\langle                 v + 2x + 2\rangle $ & $ 2^2 4^{14}$\\
$\langle v + 2x + 2\rangle $ & $\langle     2x^2 + 2x + v\rangle $ & $\langle                           1\rangle $ & $\langle                           0\rangle $ & $ 2^2 4^{14}$\\
$\langle v + 2x + 2\rangle $ & $\langle 2x^3 + 2x + v + 2\rangle $ & $\langle                           v\rangle $ & $\langle                       v + 2\rangle $ & $ 2^2 4^{14}$\\
$\langle     v + 2x\rangle $ & $\langle 2x^3 + 2x + v + 2\rangle $ & $\langle                   2x^3 + v\rangle $ & $\langle                     v + 2x\rangle $ & $ 2^2 4^{14}$\\
$\langle v + 2x + 2\rangle $ & $\langle 2x^3 + 2x + v + 2\rangle $ & $\langle                   2x^2 + v\rangle $ & $\langle         2x^2 + 2x + v + 2\rangle $ & $ 2^2 4^{14}$\\
$\langle     v + 2x\rangle $ & $\langle 2x^3 + 2x + v + 2\rangle $ & $\langle             2x^3 + 2x + v\rangle $ & $\langle     2x^3 + 2x^2 + 2x + v\rangle $ & $ 2^2 4^{14}$\\
$\langle v + 2x + 2\rangle $ & $\langle 2x^3 + 2x + v + 2\rangle $ & $\langle             2x^3 + 2x + v\rangle $ & $\langle     2x^3 + 2x^2 + 2x + v\rangle $ & $ 2^2 4^{14}$\\
$\langle     v + 2x\rangle $ & $\langle 2x^3 + 2x + v + 2\rangle $ & $\langle                       v + 2\rangle $ & $\langle                           v\rangle $ & $ 2^2 4^{14}$\\
$\langle     v + 2x\rangle $ & $\langle 2x^3 + 2x + v + 2\rangle $ & $\langle                           1\rangle $ & $\langle                           0\rangle $ & $ 2^2 4^{14}$\\
$\langle v + 2x + 2\rangle $ & $\langle 2x^3 + 2x + v + 2\rangle $ & $\langle                           1\rangle $ & $\langle                           0\rangle $ & $ 2^2 4^{14}$\\
$\langle     v + 2x\rangle $ & $\langle 2x^2 + 2x + v + 2\rangle $ & $\langle                           v\rangle $ & $\langle                       v + 2\rangle $ & $ 2^2 4^{14}$\\
$\langle     v + 2x\rangle $ & $\langle 2x^2 + 2x + v + 2\rangle $ & $\langle           2x^3 + 2x^2 + v\rangle $ & $\langle                   2x^2 + v\rangle $ & $ 2^2 4^{14}$\\
$\langle     v + 2x\rangle $ & $\langle 2x^2 + 2x + v + 2\rangle $ & $\langle     2x^3 + 2x^2 + 2x + v\rangle $ & $\langle                   2x^3 + v\rangle $ & $ 2^2 4^{14}$\\
$\langle v + 2x + 2\rangle $ & $\langle 2x^2 + 2x + v + 2\rangle $ & $\langle     2x^3 + 2x^2 + 2x + v\rangle $ & $\langle                   2x^3 + v\rangle $ & $ 2^2 4^{14}$\\
$\langle v + 2x + 2\rangle $ & $\langle 2x^2 + 2x + v + 2\rangle $ & $\langle                       v + 2\rangle $ & $\langle                           v\rangle $ & $ 2^2 4^{14}$\\
$\langle     v + 2x\rangle $ & $\langle 2x^2 + 2x + v + 2\rangle $ & $\langle                 v + 2x + 2\rangle $ & $\langle           2x^3 + 2x^2 + v\rangle $ & $ 2^2 4^{14}$\\
$\langle v + 2x + 2\rangle $ & $\langle 2x^2 + 2x + v + 2\rangle $ & $\langle                 v + 2x + 2\rangle $ & $\langle           2x^3 + 2x^2 + v\rangle $ & $ 2^2 4^{14}$\\
$\langle     v + 2x\rangle $ & $\langle 2x^2 + 2x + v + 2\rangle $ & $\langle                           1\rangle $ & $\langle                           0\rangle $ & $ 2^2 4^{14}$\\
\hline
\end{tabular}
\end{center}

There are $92$ $2$-quasi-cyclic self-dual codes over $\mathbb{Z}_4$ of length $30$ with minimum Lee weight $10$ derived from the $315$ self-dual codes over $\mathbb{Z}_4+v\mathbb{Z}_4$. These $92$ codes are given by the following table, with $C_1$, $C_2$, $C_3$, $C_4$ and the type of each $\mathbb{Z}_4$-code
$\theta(\mathcal{C})$.

\begin{center}\tiny
\begin{tabular}{llll|l}\hline
  $C_1$     & $C_2$     &     $C_3$ (mod $f_3(x)$) & $C_4$ (mod $f_4(x)$)& Type\\ \hline
$\langle           2\rangle $ & $\langle     2x^3 + 2x + v\rangle $ & $\langle                         2v\rangle $ & $\langle                       2,v\rangle $ & $ 2^{14} 4^8$\\
$\langle           2\rangle $ & $\langle     2x^3 + 2x + v\rangle $ & $\langle                       2,v\rangle $ & $\langle                         2v\rangle $ & $ 2^{14} 4^8$\\
$\langle           2\rangle $ & $\langle     2x^2 + 2x + v\rangle $ & $\langle                         2v\rangle $ & $\langle                       2,v\rangle $ & $ 2^{14} 4^8$\\
$\langle           2\rangle $ & $\langle     2x^2 + 2x + v\rangle $ & $\langle                       2,v\rangle $ & $\langle                         2v\rangle $ & $ 2^{14} 4^8$\\
$\langle           2\rangle $ & $\langle 2x^3 + 2x + v + 2\rangle $ & $\langle                         2v\rangle $ & $\langle                       2,v\rangle $ & $ 2^{14} 4^8$\\
$\langle           2\rangle $ & $\langle 2x^3 + 2x + v + 2\rangle $ & $\langle                       2,v\rangle $ & $\langle                         2v\rangle $ & $ 2^{14} 4^8$\\
$\langle           2\rangle $ & $\langle 2x^2 + 2x + v + 2\rangle $ & $\langle                         2v\rangle $ & $\langle                       2,v\rangle $ & $ 2^{14} 4^8$\\
$\langle           2\rangle $ & $\langle 2x^2 + 2x + v + 2\rangle $ & $\langle                       2,v\rangle $ & $\langle                         2v\rangle $ & $ 2^{14} 4^8$\\
$\langle     v + 2x\rangle $ & $\langle     2x^3 + 2x + v\rangle $ & $\langle                         2v\rangle $ & $\langle                       2,v\rangle $ & $ 2^{10} 4^{10}$\\
$\langle v + 2x + 2\rangle $ & $\langle     2x^3 + 2x + v\rangle $ & $\langle                         2v\rangle $ & $\langle                       2,v\rangle $ & $ 2^{10} 4^{10}$\\
$\langle v + 2x + 2\rangle $ & $\langle     2x^3 + 2x + v\rangle $ & $\langle                       2,v\rangle $ & $\langle                         2v\rangle $ & $ 2^{10} 4^{10}$\\
$\langle v + 2x + 2\rangle $ & $\langle     2x^2 + 2x + v\rangle $ & $\langle                         2v\rangle $ & $\langle                       2,v\rangle $ & $ 2^{10} 4^{10}$\\
$\langle     v + 2x\rangle $ & $\langle     2x^2 + 2x + v\rangle $ & $\langle                       2,v\rangle $ & $\langle                         2v\rangle $ & $ 2^{10} 4^{10}$\\
$\langle v + 2x + 2\rangle $ & $\langle     2x^2 + 2x + v\rangle $ & $\langle                       2,v\rangle $ & $\langle                         2v\rangle $ & $ 2^{10} 4^{10}$\\
$\langle     v + 2x\rangle $ & $\langle 2x^3 + 2x + v + 2\rangle $ & $\langle                         2v\rangle $ & $\langle                       2,v\rangle $ & $ 2^{10} 4^{10}$\\
$\langle v + 2x + 2\rangle $ & $\langle 2x^3 + 2x + v + 2\rangle $ & $\langle                         2v\rangle $ & $\langle                       2,v\rangle $ & $ 2^{10} 4^{10}$\\
$\langle v + 2x + 2\rangle $ & $\langle 2x^3 + 2x + v + 2\rangle $ & $\langle                       2,v\rangle $ & $\langle                         2v\rangle $ & $ 2^{10} 4^{10}$\\
$\langle v + 2x + 2\rangle $ & $\langle 2x^2 + 2x + v + 2\rangle $ & $\langle                         2v\rangle $ & $\langle                       2,v\rangle $ & $ 2^{10} 4^{10}$\\
$\langle     v + 2x\rangle $ & $\langle 2x^2 + 2x + v + 2\rangle $ & $\langle                       2,v\rangle $ & $\langle                         2v\rangle $ & $ 2^{10} 4^{10}$\\
$\langle v + 2x + 2\rangle $ & $\langle 2x^2 + 2x + v + 2\rangle $ & $\langle                       2,v\rangle $ & $\langle                         2v\rangle $ & $ 2^{10} 4^{10}$\\
$\langle     v + 2x\rangle $ & $\langle                   2\rangle $ & $\langle                   2x^2 + v\rangle $ & $\langle         2x^2 + 2x + v + 2\rangle $ & $ 2^{10} 4^{10}$\\
$\langle v + 2x + 2\rangle $ & $\langle                   2\rangle $ & $\langle                   2x^2 + v\rangle $ & $\langle         2x^2 + 2x + v + 2\rangle $ & $ 2^{10} 4^{10}$\\
$\langle     v + 2x\rangle $ & $\langle                   2\rangle $ & $\langle                     v + 2x\rangle $ & $\langle       2x^3 + 2x^2 + v + 2\rangle $ & $ 2^{10} 4^{10}$\\
$\langle v + 2x + 2\rangle $ & $\langle                   2\rangle $ & $\langle                     v + 2x\rangle $ & $\langle       2x^3 + 2x^2 + v + 2\rangle $ & $ 2^{10} 4^{10}$\\
$\langle     v + 2x\rangle $ & $\langle                   2\rangle $ & $\langle               2x^2 + v + 2\rangle $ & $\langle             2x^2 + 2x + v\rangle $ & $ 2^{10} 4^{10}$\\
$\langle v + 2x + 2\rangle $ & $\langle                   2\rangle $ & $\langle               2x^2 + v + 2\rangle $ & $\langle             2x^2 + 2x + v\rangle $ & $ 2^{10} 4^{10}$\\
$\langle     v + 2x\rangle $ & $\langle                   2\rangle $ & $\langle                 v + 2x + 2\rangle $ & $\langle           2x^3 + 2x^2 + v\rangle $ & $ 2^{10} 4^{10}$\\
$\langle v + 2x + 2\rangle $ & $\langle                   2\rangle $ & $\langle                 v + 2x + 2\rangle $ & $\langle           2x^3 + 2x^2 + v\rangle $ & $ 2^{10} 4^{10}$\\
$\langle           2\rangle $ & $\langle     2x^3 + 2x + v\rangle $ & $\langle                           v\rangle $ & $\langle                       v + 2\rangle $ & $ 2^6 4^{12}$\\
$\langle           2\rangle $ & $\langle     2x^3 + 2x + v\rangle $ & $\langle           2x^3 + 2x^2 + v\rangle $ & $\langle                   2x^2 + v\rangle $ & $ 2^6 4^{12}$\\
$\langle           2\rangle $ & $\langle     2x^3 + 2x + v\rangle $ & $\langle             2x^3 + 2x + v\rangle $ & $\langle     2x^3 + 2x^2 + 2x + v\rangle $ & $ 2^6 4^{12}$\\
$\langle           2\rangle $ & $\langle     2x^3 + 2x + v\rangle $ & $\langle                       v + 2\rangle $ & $\langle                           v\rangle $ & $ 2^6 4^{12}$\\
$\langle           2\rangle $ & $\langle     2x^3 + 2x + v\rangle $ & $\langle       2x^3 + 2x^2 + v + 2\rangle $ & $\langle               2x^2 + v + 2\rangle $ & $ 2^6 4^{12}$\\
$\langle           2\rangle $ & $\langle     2x^3 + 2x + v\rangle $ & $\langle         2x^3 + 2x + v + 2\rangle $ & $\langle 2x^3 + 2x^2 + 2x + v + 2\rangle $ & $ 2^6 4^{12}$\\
$\langle           2\rangle $ & $\langle     2x^3 + 2x + v\rangle $ & $\langle                           0\rangle $ & $\langle                           1\rangle $ & $ 2^6 4^{12}$\\
$\langle           2\rangle $ & $\langle     2x^2 + 2x + v\rangle $ & $\langle                           v\rangle $ & $\langle                       v + 2\rangle $ & $ 2^6 4^{12}$\\
$\langle           2\rangle $ & $\langle     2x^2 + 2x + v\rangle $ & $\langle                   2x^3 + v\rangle $ & $\langle                     v + 2x\rangle $ & $ 2^6 4^{12}$\\
$\langle           2\rangle $ & $\langle     2x^2 + 2x + v\rangle $ & $\langle     2x^3 + 2x^2 + 2x + v\rangle $ & $\langle                   2x^3 + v\rangle $ & $ 2^6 4^{12}$\\
$\langle           2\rangle $ & $\langle     2x^2 + 2x + v\rangle $ & $\langle                       v + 2\rangle $ & $\langle                           v\rangle $ & $ 2^6 4^{12}$\\
$\langle           2\rangle $ & $\langle     2x^2 + 2x + v\rangle $ & $\langle               2x^3 + v + 2\rangle $ & $\langle                 v + 2x + 2\rangle $ & $ 2^6 4^{12}$\\
$\langle           2\rangle $ & $\langle     2x^2 + 2x + v\rangle $ & $\langle 2x^3 + 2x^2 + 2x + v + 2\rangle $ & $\langle               2x^3 + v + 2\rangle $ & $ 2^6 4^{12}$\\
$\langle           2\rangle $ & $\langle 2x^3 + 2x + v + 2\rangle $ & $\langle                           v\rangle $ & $\langle                       v + 2\rangle $ & $ 2^6 4^{12}$\\
$\langle           2\rangle $ & $\langle 2x^3 + 2x + v + 2\rangle $ & $\langle           2x^3 + 2x^2 + v\rangle $ & $\langle                   2x^2 + v\rangle $ & $ 2^6 4^{12}$\\
$\langle           2\rangle $ & $\langle 2x^3 + 2x + v + 2\rangle $ & $\langle             2x^3 + 2x + v\rangle $ & $\langle     2x^3 + 2x^2 + 2x + v\rangle $ & $ 2^6 4^{12}$\\
$\langle           2\rangle $ & $\langle 2x^3 + 2x + v + 2\rangle $ & $\langle                       v + 2\rangle $ & $\langle                           v\rangle $ & $ 2^6 4^{12}$\\
$\langle           2\rangle $ & $\langle 2x^3 + 2x + v + 2\rangle $ & $\langle       2x^3 + 2x^2 + v + 2\rangle $ & $\langle               2x^2 + v + 2\rangle $ & $ 2^6 4^{12}$\\
$\langle           2\rangle $ & $\langle 2x^3 + 2x + v + 2\rangle $ & $\langle         2x^3 + 2x + v + 2\rangle $ & $\langle 2x^3 + 2x^2 + 2x + v + 2\rangle $ & $ 2^6 4^{12}$\\
$\langle           2\rangle $ & $\langle 2x^3 + 2x + v + 2\rangle $ & $\langle                           0\rangle $ & $\langle                           1\rangle $ & $ 2^6 4^{12}$\\
$\langle           2\rangle $ & $\langle 2x^2 + 2x + v + 2\rangle $ & $\langle                           v\rangle $ & $\langle                       v + 2\rangle $ & $ 2^6 4^{12}$\\
$\langle           2\rangle $ & $\langle 2x^2 + 2x + v + 2\rangle $ & $\langle                   2x^3 + v\rangle $ & $\langle                     v + 2x\rangle $ & $ 2^6 4^{12}$\\
$\langle           2\rangle $ & $\langle 2x^2 + 2x + v + 2\rangle $ & $\langle     2x^3 + 2x^2 + 2x + v\rangle $ & $\langle                   2x^3 + v\rangle $ & $ 2^6 4^{12}$\\
$\langle           2\rangle $ & $\langle 2x^2 + 2x + v + 2\rangle $ & $\langle                       v + 2\rangle $ & $\langle                           v\rangle $ & $ 2^6 4^{12}$\\
$\langle           2\rangle $ & $\langle 2x^2 + 2x + v + 2\rangle $ & $\langle               2x^3 + v + 2\rangle $ & $\langle                 v + 2x + 2\rangle $ & $ 2^6 4^{12}$\\
$\langle           2\rangle $ & $\langle 2x^2 + 2x + v + 2\rangle $ & $\langle 2x^3 + 2x^2 + 2x + v + 2\rangle $ & $\langle               2x^3 + v + 2\rangle $ & $ 2^6 4^{12}$\\
$\langle     v + 2x\rangle $ & $\langle     2x^3 + 2x + v\rangle $ & $\langle                           v\rangle $ & $\langle                       v + 2\rangle $ & $ 2^2 4^{14}$\\
$\langle     v + 2x\rangle $ & $\langle     2x^3 + 2x + v\rangle $ & $\langle             2x^3 + 2x + v\rangle $ & $\langle     2x^3 + 2x^2 + 2x + v\rangle $ & $ 2^2 4^{14}$\\
$\langle v + 2x + 2\rangle $ & $\langle     2x^3 + 2x + v\rangle $ & $\langle             2x^3 + 2x + v\rangle $ & $\langle     2x^3 + 2x^2 + 2x + v\rangle $ & $ 2^2 4^{14}$\\
$\langle v + 2x + 2\rangle $ & $\langle     2x^3 + 2x + v\rangle $ & $\langle                       v + 2\rangle $ & $\langle                           v\rangle $ & $ 2^2 4^{14}$\\
$\langle     v + 2x\rangle $ & $\langle     2x^3 + 2x + v\rangle $ & $\langle               2x^2 + v + 2\rangle $ & $\langle             2x^2 + 2x + v\rangle $ & $ 2^2 4^{14}$\\
$\langle     v + 2x\rangle $ & $\langle     2x^3 + 2x + v\rangle $ & $\langle                 v + 2x + 2\rangle $ & $\langle           2x^3 + 2x^2 + v\rangle $ & $ 2^2 4^{14}$\\
$\langle v + 2x + 2\rangle $ & $\langle     2x^3 + 2x + v\rangle $ & $\langle                 v + 2x + 2\rangle $ & $\langle           2x^3 + 2x^2 + v\rangle $ & $ 2^2 4^{14}$\\
$\langle     v + 2x\rangle $ & $\langle     2x^3 + 2x + v\rangle $ & $\langle                           0\rangle $ & $\langle                           1\rangle $ & $ 2^2 4^{14}$\\
$\langle v + 2x + 2\rangle $ & $\langle     2x^3 + 2x + v\rangle $ & $\langle                           0\rangle $ & $\langle                           1\rangle $ & $ 2^2 4^{14}$\\
$\langle v + 2x + 2\rangle $ & $\langle     2x^2 + 2x + v\rangle $ & $\langle                           v\rangle $ & $\langle                       v + 2\rangle $ & $ 2^2 4^{14}$\\
$\langle     v + 2x\rangle $ & $\langle     2x^2 + 2x + v\rangle $ & $\langle     2x^3 + 2x^2 + 2x + v\rangle $ & $\langle                   2x^3 + v\rangle $ & $ 2^2 4^{14}$\\
$\langle v + 2x + 2\rangle $ & $\langle     2x^2 + 2x + v\rangle $ & $\langle     2x^3 + 2x^2 + 2x + v\rangle $ & $\langle                   2x^3 + v\rangle $ & $ 2^2 4^{14}$\\
$\langle     v + 2x\rangle $ & $\langle     2x^2 + 2x + v\rangle $ & $\langle                       v + 2\rangle $ & $\langle                           v\rangle $ & $ 2^2 4^{14}$\\
$\langle     v + 2x\rangle $ & $\langle     2x^2 + 2x + v\rangle $ & $\langle               2x^2 + v + 2\rangle $ & $\langle             2x^2 + 2x + v\rangle $ & $ 2^2 4^{14}$\\
$\langle v + 2x + 2\rangle $ & $\langle     2x^2 + 2x + v\rangle $ & $\langle               2x^2 + v + 2\rangle $ & $\langle             2x^2 + 2x + v\rangle $ & $ 2^2 4^{14}$\\
$\langle     v + 2x\rangle $ & $\langle     2x^2 + 2x + v\rangle $ & $\langle 2x^3 + 2x^2 + 2x + v + 2\rangle $ & $\langle               2x^3 + v + 2\rangle $ & $ 2^2 4^{14}$\\
$\langle     v + 2x\rangle $ & $\langle     2x^2 + 2x + v\rangle $ & $\langle                           1\rangle $ & $\langle                           0\rangle $ & $ 2^2 4^{14}$\\
$\langle     v + 2x\rangle $ & $\langle     2x^2 + 2x + v\rangle $ & $\langle                           0\rangle $ & $\langle                           1\rangle $ & $ 2^2 4^{14}$\\
$\langle v + 2x + 2\rangle $ & $\langle     2x^2 + 2x + v\rangle $ & $\langle                           0\rangle $ & $\langle                           1\rangle $ & $ 2^2 4^{14}$\\
$\langle     v + 2x\rangle $ & $\langle 2x^3 + 2x + v + 2\rangle $ & $\langle                           v\rangle $ & $\langle                       v + 2\rangle $ & $ 2^2 4^{14}$\\
$\langle     v + 2x\rangle $ & $\langle 2x^3 + 2x + v + 2\rangle $ & $\langle                   2x^2 + v\rangle $ & $\langle         2x^2 + 2x + v + 2\rangle $ & $ 2^2 4^{14}$\\
$\langle     v + 2x\rangle $ & $\langle 2x^3 + 2x + v + 2\rangle $ & $\langle           2x^3 + 2x^2 + v\rangle $ & $\langle                   2x^2 + v\rangle $ & $ 2^2 4^{14}$\\
$\langle v + 2x + 2\rangle $ & $\langle 2x^3 + 2x + v + 2\rangle $ & $\langle           2x^3 + 2x^2 + v\rangle $ & $\langle                   2x^2 + v\rangle $ & $ 2^2 4^{14}$\\
$\langle     v + 2x\rangle $ & $\langle 2x^3 + 2x + v + 2\rangle $ & $\langle                     v + 2x\rangle $ & $\langle       2x^3 + 2x^2 + v + 2\rangle $ & $ 2^2 4^{14}$\\
$\langle v + 2x + 2\rangle $ & $\langle 2x^3 + 2x + v + 2\rangle $ & $\langle                     v + 2x\rangle $ & $\langle       2x^3 + 2x^2 + v + 2\rangle $ & $ 2^2 4^{14}$\\
$\langle v + 2x + 2\rangle $ & $\langle 2x^3 + 2x + v + 2\rangle $ & $\langle                       v + 2\rangle $ & $\langle                           v\rangle $ & $ 2^2 4^{14}$\\
$\langle     v + 2x\rangle $ & $\langle 2x^3 + 2x + v + 2\rangle $ & $\langle                           0\rangle $ & $\langle                           1\rangle $ & $ 2^2 4^{14}$\\
$\langle v + 2x + 2\rangle $ & $\langle 2x^3 + 2x + v + 2\rangle $ & $\langle                           0\rangle $ & $\langle                           1\rangle $ & $ 2^2 4^{14}$\\
$\langle v + 2x + 2\rangle $ & $\langle 2x^2 + 2x + v + 2\rangle $ & $\langle                           v\rangle $ & $\langle                       v + 2\rangle $ & $ 2^2 4^{14}$\\
$\langle     v + 2x\rangle $ & $\langle 2x^2 + 2x + v + 2\rangle $ & $\langle                   2x^3 + v\rangle $ & $\langle                     v + 2x\rangle $ & $ 2^2 4^{14}$\\
$\langle     v + 2x\rangle $ & $\langle 2x^2 + 2x + v + 2\rangle $ & $\langle                   2x^2 + v\rangle $ & $\langle         2x^2 + 2x + v + 2\rangle $ & $ 2^2 4^{14}$\\
$\langle v + 2x + 2\rangle $ & $\langle 2x^2 + 2x + v + 2\rangle $ & $\langle                   2x^2 + v\rangle $ & $\langle         2x^2 + 2x + v + 2\rangle $ & $ 2^2 4^{14}$\\
$\langle     v + 2x\rangle $ & $\langle 2x^2 + 2x + v + 2\rangle $ & $\langle                       v + 2\rangle $ & $\langle                           v\rangle $ & $ 2^2 4^{14}$\\
$\langle     v + 2x\rangle $ & $\langle 2x^2 + 2x + v + 2\rangle $ & $\langle               2x^3 + v + 2\rangle $ & $\langle                 v + 2x + 2\rangle $ & $ 2^2 4^{14}$\\
$\langle v + 2x + 2\rangle $ & $\langle 2x^2 + 2x + v + 2\rangle $ & $\langle               2x^3 + v + 2\rangle $ & $\langle                 v + 2x + 2\rangle $ & $ 2^2 4^{14}$\\
$\langle v + 2x + 2\rangle $ & $\langle 2x^2 + 2x + v + 2\rangle $ & $\langle                           1\rangle $ & $\langle                           0\rangle $ & $ 2^2 4^{14}$\\
$\langle     v + 2x\rangle $ & $\langle 2x^2 + 2x + v + 2\rangle $ & $\langle                           0\rangle $ & $\langle                           1\rangle $ & $ 2^2 4^{14}$\\
$\langle v + 2x + 2\rangle $ & $\langle 2x^2 + 2x + v + 2\rangle $ & $\langle                           0\rangle $ & $\langle                           1\rangle $ & $ 2^2 4^{14}$\\
\hline
\end{tabular}
\end{center}

\vskip 3mm\par
  Let $\phi$ be the Gray map from $\mathbb{Z}_4^{30}$ onto $\mathbb{F}_2^{60}$ extended by $0\rightarrow 00, 1\rightarrow 01,
2\rightarrow 11, 3\rightarrow 10$ in the natural way. Then $\phi$ is a distance and orthogonality preserving bijection
from $(\mathbb{Z}_4^{30}, {\rm Lee} \ {\rm distance})$ onto $(\mathbb{F}_2^{60}, {\rm Hamming} \ {\rm distance})$.
Let $D$ be any self-dual linear code over $\mathbb{Z}_4$ of length $30$. Then $D=D^\bot$. From $|D|=|D^\bot|$ and
$[\xi,\eta]=0$ for all $\xi\in D$ and $\eta\in D^\bot$, we deduce that
$|\phi(D)|=|\phi(D^\bot)|$ and
$[\phi(\xi),\phi(\eta)]=0$ for all $\phi(\xi)\in \phi(D)$ and $\phi(\eta)\in \phi(D^\bot)$. This implies
$\phi(D)^\bot=\phi(D^\bot)=\phi(D)$, i.e. $\phi(D)$ is a binary self-dual code of length $60$. Let
$\alpha,\beta\in \phi(D)$. For any  $\gamma\in \phi(D)$, by $[\alpha,\gamma]=[\beta,\gamma]=0$ it follows that $[\alpha+\beta,\gamma]=[\alpha,\gamma]+[\beta,\gamma]=0$. This implies
$\alpha+\beta\in \phi(D)^\bot=\phi(D)$. Hence $\phi(D)$ is a binary linear code self-dual code of length $60$.

\vskip 3mm\noindent
 {\bf Remark} From the $70$ $2$-quasi-cyclic self-dual codes with minimal Lee weight $12$
and $92$ $2$-quasi-cyclic self-dual codes with minimum Lee weight $10$ over $\mathbb{Z}_4$ listed above and by the Gray map $\phi$, we derive $70$ $4$-quasi-cyclic binary self-dual $[60,30,12]$ codes
and $92$ $4$-quasi-cyclic binary self-dual $[60,30,10]$ codes. It is well known that
binary self-dual $[60,30,12]$ codes are extremal.

\vskip 3mm\par
  Recently, Shi et al. in [15] studied $(1+2u)$-constacyclic codes over the ring
$\Gamma=\mathbb{Z}_{4}[u]/\langle u^2-1\rangle=\mathbb{Z}_{4}+u\mathbb{Z}_{4}$ ($u^2=1$) of odd length $n$, i.e. ideals of the ring $\Gamma[x]/\langle x^n-(1+2u)\rangle$. Properties of these codes and
their $\mathbb{Z}_{4}$ images were investigated.

\par
  Define a map $\varrho$ by: for any $\gamma(x)=\sum_{i=0}^{n-1}\gamma_ix^i\in \Gamma[x]/\langle x^n-1\rangle$,
$$\varrho(\gamma(x))=\gamma((1+2u)x)=\sum_{i=0}^{n-1}\gamma_i(1+2u)^ix^i.$$
Then $\varrho$ is an isomorphism of rings from $\Gamma[x]/\langle x^n-1\rangle$ onto $\Gamma[x]/\langle x^n-(1+2u)\rangle$ ([15] Proposition 3.6) (and a Hamming distance-preserving map over $\Gamma$). Hence $\mathcal{E}$ is a $(1+2u)$-constacyclic code over
$\Gamma$ of length $n$ if and only if there is a unique cyclic code $\mathcal{D}$ over
$\Gamma$ of length $n$ such that $\varrho(\mathcal{D})=\mathcal{E}$ (cf. [15] Lemma 3.9).

\par
  Now, we set $v=u-1$. By $(u-1)^2=2(u-1)$ in
$\Gamma$, it follows that
$$\varsigma: a+bv\mapsto a+b(u-1)=(a-b)+bu, \ \forall a,b\in \mathbb{Z}_{4}$$
is an isomorphism of rings
from $R=\mathbb{Z}_{4}[v]/\langle v^2+2v\rangle$ onto $\Gamma=\mathbb{Z}_{4}[u]/\langle u^2-1\rangle$. Then $\varsigma$ induces a natural ring isomorphism
from $\frac{(\mathbb{Z}_{4}[v]/\langle v^2+2v\rangle)[x]}{\langle x^n-1\rangle}$ onto
$\frac{(\mathbb{Z}_{4}[u]/\langle u^2-1\rangle)[x]}{\langle x^n-1\rangle}$ by
$$\sum_{i=0}^{n-1}(a_i+b_iv)x^i\mapsto \sum_{i=0}^{n-1}\left((a_i-b_i)+b_iu\right)x^i,
\ \forall a_i,b_i\in \mathbb{Z}_{4}, 0\leq i\leq n-1.$$
From this, we deduce the following conclusion.

\vskip 3mm\noindent
  {\bf Proposition 5.4} \textit{Using the notations above, let $n$ be an odd positive integer. Then $\mathcal{E}$ is a cyclic code over
$\Gamma=\mathbb{Z}_{4}[u]/\langle u^2-1\rangle$ of length $n$ if and only if there is a unique cyclic code $\mathcal{C}$ over
$R=\mathbb{Z}_{4}[v]/\langle v^2+2v\rangle$ of length $n$ such that}
$$\mathcal{E}=\left\{\sum_{i=0}^{n-1}\left((a_i-b_i)+b_iu\right)x^i\mid a_i,b_i\in \mathbb{Z}_{4}, \ \sum_{i=0}^{n-1}(a_i+b_iv)x^i\in \mathcal{C}\right\}.$$

\vskip 3mm\par
  Therefore, in order to give a complete classification for all distinct
cyclic codes over
$\Gamma=\mathbb{Z}_{4}[u]/\langle u^2-1\rangle$ of odd length $n$, we only need to present
all distinct cyclic codes $\mathcal{C}$ over
$R=\mathbb{Z}_{4}[v]/\langle v^2+2v\rangle$ of length $n$.

\begin{acknowledgements}
Part of this work was done when Yonglin Cao was visiting Chern Institute of Mathematics, Nankai University, Tianjin, China. Yonglin Cao would like to thank the institution for the kind hospitality. This research is
supported in part by the National Natural Science Foundation of
China (Grant Nos. 11671235, 11471255).
\end{acknowledgements}

\vskip 5mm \noindent
{\bf Appendix: Proof of Theorem 2.3}

\vskip 3mm
   Using the notations of Section 2, by Lemma 2.1, [6] Example 2.5 we know that the number of
linear codes over the Galois ring $A_f$ of length $2$ is equal to
\begin{center}
$\sum_{i=0}^s(2i+1)|T_f|^{s-i}=\sum_{i=0}^s(2i+1)p^{(s-i)m}$.
\end{center}
Moreover, every nontrivial linear code $C$ over
$A_f$ of length $2$ has one and only one of the following matrices $G$ as their generator matrices:

\vskip 2mm \par
 (i) \textit{$G=(1,a)$, $a\in A_f$}.

\vskip 2mm \par
 (ii) \textit{$G=(p^k,p^ka)$, $a\in A_f/\langle p^{s-k}\rangle$, $1\leq k\leq s-1$}.

\vskip 2mm \par
(iii) \textit{$G=(p b,1)$, $b\in A_f/\langle p^{s-1}\rangle$}.

\vskip 2mm \par
(iv) \textit{$G=(p^{k+1}b,p^k)$, $b\in A_f/\langle p^{s-k-1}\rangle$, $1\leq k\leq s-1$}.

\vskip 2mm \par
 (v) \textit{$G=p^kI_2$, $1\leq k\leq s-1$}.

\vskip 2mm \par
  (vi) \textit{$G=\left(\begin{array}{cc}1 & c\cr
0 & p^t\end{array}\right)$,  $c\in A_f/\langle p^{t}\rangle$, $1\leq t\leq s-1$}.

\vskip 2mm \par
  (vii) \textit{$G=\left(\begin{array}{cc} p^k & p^kc\cr
0 & p^{k+t}\end{array}\right)$,  $c\in A_f/\langle p^{t}\rangle$, $1\leq t\leq s-k-1$, $1\leq k\leq s-2$}.

\vskip 2mm \par
    (viii) \textit{$G=\left(\begin{array}{cc}0 & 1\cr p^t & 0\end{array}\right)$ if $s=2$; $G=\left(\begin{array}{cc}c & 1\cr p^t & 0\end{array}\right)$, where $c\in p(A_f/\langle p^{t}\rangle)$ and $1\leq t\leq s-1$, if $s\geq 3$}.

\vskip 2mm \par
    (ix) \textit{$G=\left(\begin{array}{cc}p^kc & p^k\cr p^{k+t} & 0\end{array}\right)$, $c\in p(A_f/\langle p^{t}\rangle)$,
$1\leq t\leq s-k-1$, $1\leq k\leq s-2$}.

\vskip 3mm \noindent
Therefore, we only need to consider the nine cases listed above:

\vskip 2mm\par
   (i) Suppose that $C$ satisfies Condition (2). By $(1,a)\in C$, we have
$(0,1+pa)\in C$. Since $G$ is the generator matrix of $C$, there exists $b\in A_f$ such that $(0,1+pa)=b(1,a)=(b,ba)$,
i.e. $0=b$ and $1+pa=ba$, which implies $1+pa=0$. Hence we get a contradiction, since $1+pa$ is an invertible element
of $A_f$ by Lemma 2.1. Therefore, $C$ does not satisfy Condition (2) in this case.

\par
   (ii) Suppose that $C$ satisfies Condition (2). By $(p^k,p^ka)\in C$,
we have $(0,p^k+p^{k+1}a)=(0,p^k+p\cdot p^ka)\in C$. Then there exists $b\in A_f$ such that $(0,p^k+p^{k+1}a)=b(p^k,p^k a)=(p^k b,p^k ba)$,
which implies $0=p^k b$ and $p^k(1+pa) =\pi^k ba$, and so $p^k(1+pa) =0$. Since $1+pa$ is an invertible element
of $A_f$, we deduce $p^k=0$, which contradict that $k\leq s-1$.

\par
   (iii) In this case, $C$ satisfies Condition (2) if and only if
there exists $a\in A_f$ such that $(0,pb+p)=(0,pb+p\cdot 1)=a(p b,1)=(p ab,a)$, i.e., $0=p ab$ and $p(b+1)=pb+p=a$. These conditions
are equivalent to that $b$ satisfies $p^2(b+1)b=0$, i.e. $(b+1)b\in p^{s-2}A_f$. Then we have one of the following two cases:

\par
  (iii-1) If $s=2$, then $(b+1)b\in A_f=p^{s-2}A_f$ for any $b\in A_f/\langle p^{s-1}\rangle=A_f/\langle p\rangle=T_f$.

\par
  (iii-2) Let $s\geq 2$. Then $(b+1)b\in p^{s-2}A_f$ is equivalent to $\|(b+1)b\|_p\geq s-2$.
  By Lemma 2.1(iv) and $b\in A_f/\langle p^{s-1}\rangle$, $b$ has a unique $p$-expansion:
$b=\sum_{i=0}^{s-2}t_ip^i$ with $t_0,t_1,\ldots,t_{s-2}\in T_f$, which implies $b+1=t_0+1+\sum_{i=1}^{s-2}t_ip^i$.
Then we have one of the following subcases:

\par
  (iii-2-1) When $t_0+1\neq 0$, i.e. $t_0\neq 1$, $b+1$ is invertible in $A_f$. From this we deduce that $\|(b+1)b\|_p=\|b\|_p$. In this case,
$\|(b+1)b\|_p\geq s-2$ if and only if $b=t_{s-2}p^{s-2}$ with $t_{s-2}\in T_f$.

\par
  (iii-2-2) When $t_0+1=0$, i.e., $t_0=-1$, $b$ is invertible in $A_f$. From this we deduce that $\|(b+1)b\|_p=\|b+1\|_p$. In this case,
$\|(b+1)b\|_p\geq s-2$ if and only if $b+1=t_{s-2}p^{s-2}$, i.e. $b=-1+t_{s-2}p^{s-2}$, where $t_{s-2}\in T_f$.

\par
   (iv) In this case, $C$ satisfies Condition (2) if and only if
there exists $a\in A_f$ such that $(0,p^{k+1}(b+1))=(0,p^{k+1}b+p\cdot p^{k})=a(p^{k+1}b,p^k)=(p^{k+1}ab,p^ka)$, i.e. $0=p^{k+1}ab=pb\cdot p^ka$ and $p^{k+1}(b+1)=p^ka$. These conditions
are equivalent to that $b$ satisfies $p^{k+2}(b+1)b=0$, i.e., $\|(b+1)b\|_p\geq s-k-2$.
As $1\leq k\leq s-1$, we have $1\leq s-k\leq s-1$, which implies
$-1\leq s-k-2\leq s-3$.
Then we have one of the following three cases:

\par
  (iv-1) If $s=2$, then $\|(b+1)b\|_p\geq s-k-2=-k$ for any $b\in A_f/\langle p^{s-k-1}\rangle=A_f/\langle p^{0}\rangle
=\{0\}$. In this case, we have $G=(0,p)$.

\par
  (iv-2) If $s=3$, then $\|(b+1)b\|_p\geq s-k-2=1-k$ for any $b\in A_f/\langle p^{2-k}\rangle$, where $1\leq k\leq 2$.

\par
  (iv-3) Let $s\geq 4$. If $k=s-1$, then $\|(b+1)b\|_p\geq s-k-2=-1$ for any $b\in A_f//\langle p^{0}\rangle=\{0\}$ and
hence $G=(0,p^{s-1})$. If $k=s-2$, then $\|(b+1)b\|_p\geq s-k-2=0$ if and only if $b\in T_f$. In this case, we have
$G=(p^{s-1}b,p^{s-2})$.

\par
  Now, we assume $1\leq k\leq s-3$.
  By Lemma 2.1(iv) and $b\in A_f/\langle p^{s-k-1}\rangle$, $b$ has a unique $p$-expansion:
$b=\sum_{i=0}^{s-k-2}t_ip^i$ with $t_0,t_1,\ldots,t_{s-k-2}\in T_f$, which implies $b+1=t_0+1+\sum_{i=1}^{s-k-2}t_ip^i$.
Then we have one of the following subcases:

\par
  (iv-3-1) When $t_0+1\neq 0$, $b+1$ is invertible in $A_f$. From this we deduce that $\|(b+1)b\|_p=\|b\|_p$. In this case,
$\|(b+1)b\|_p\geq s-k-2$ if and only if $b=t_{s-k-2}p^{s-k-2}$ with $t_{s-k-2}\in T_f$.

\par
  (iv-3-2) When $t_0+1=0$, $b$ is invertible in $A_f$. From this we deduce that $\|(b+1)b\|_p=\|b+1\|_p$. In this case,
$\|(b+1)b\|_p\geq s-k-2$ if and only if $b+1=t_{s-k-2}p^{s-k-2}$, i.e. $b=-1+t_{s-k-2}p^{s-k-2}$, where $t_{s-k-2}\in T_f$.

\par
   (v) In this case, $C$ satisfies Condition (2) for all $1\leq k\leq s-1$.

\par
   (vi) Suppose
that $C$ satisfies Condition (2). Then there exist $a,b\in A_f$ such that $(0,1+pc)=a(1,c)+b(0,p^t)=(a,ac+p^t b)$,
i.e. $0=a$ and $1+pc=ac+p^t b$. This implies $1=p(p^{t-1} b-c)$, and we get a contradiction.
Hence $C$ does not satisfy Condition (2) in this case.

\par
   (vii) Suppose
that $C$ satisfies Condition (2). Then there exist $a,b\in A_f$ such that $(0,p^k+p\cdot p^kc)=a(p^k,p^kc)+b(0,p^{k+t})
=(p^ka,p^kac+p^{k+t}b)$,
i.e. $0=p^ka$ and $p^k+p^{k+1}c=p^kac+p^{k+t} b$. This implies $p^k=p^{k+1}(p^{t-1}b-c)$, and we get a contradiction
as $1\leq k\leq s-1$.

\par
   (viii) It is clear that $(0,p^t)=p^t(c,1)-c(p^t,0)\in C$.
Hence $C$ satisfies Condition (2) if and only if there exist $a,b\in A_f$ such that
$(0,c+p\cdot 1)=a(c,1)+b(p^t,0)=(ac+p^tb,a)$,
i.e. $0=ac+p^tb$ and $c+p=a$, which are equivalent to
that $(c+p)c=-p^tb\in p^tA_f$ for some $b\in A_f$, i.e. $\|(c+p)c\|_p\geq t$, where $c\in p(A_f/\langle p^t\rangle)$ and $1\leq t\leq s-1$.
Then we have one of the following three subcases:

\par
   (viii-1) When $t=1$, then $c\in p(A_f/\langle p\rangle)=\{0\}$, i.e. $c=0$. In this case, $\|(c+p)c\|_p=\|0\|_p=s>t$.

\par
   (viii-2) When $t\geq 2$ and $s\geq 3$, by $c\in p(A/\langle p^t\rangle)$ and Lemma 2.1(i) it follows that
$c=e_1p+\ldots+e_{t-1}p^{t-1}$ where $e_1,\ldots,e_{t-1}\in T_f$.
Then $c+p=(e_1+1)p+\ldots+e_{t-1}p^{t-1}$. If $t=2$, it is obvious that $\|(c+p)c\|_p\geq 2=t$ for all
$c\in p(A/\langle p^2\rangle)=\{e_1p\mid e_1\in T_f\}$. Now, assume $t\geq 3$ and $s\geq 4$.
Suppose that $e_1\neq 0$ and $e_1\neq -1$. Then $\|c\|_p=\|c+p\|_p=1$ and $\|(c+p)c\|_p=2<t$. Hence
By $\|(c+p)c\|_p\geq t\geq 3$, we have
one of the following two subcases:

\par
  $\triangleright$ $e_1\neq 0$ and $e_1=-1$. In this case, $\|c\|_p=1$. Then $\|(c+p)c\|_p\geq t$ if and only if
$\|c+p\|_p\geq t-1$, which is equivalent to that $e_1=-1$ and $c+p=e_{t-1}p^{t-1}$, i.e. $c=-p+e_{t-1}p^{t-1}$,
where $e_{t-1}\in T_f$.

\par
  $\triangleright$ $e_1\neq -1$ and $e_1=0$. In this case, $\|c+p\|_p=1$. Then $\|(c+p)c\|_p\geq t$ if and only if
$\|c\|_p\geq t-1$, which is equivalent to that $c=e_{t-1}p^{t-1}$ where $e_{t-1}\in T_f$.

\par
   (xi) It is clear that $(0,p^{k+t})=p^t(p^kc,p^k)-c(p^{k+t},0)\in C$.
Hence $C$ satisfies Condition (2) if and only if there exist $a,b\in A_f$ such that
$(0,p^kc+p\cdot p^k)=a(p^kc,p^k)+b(p^{k+t},0)=(p^kac+p^{k+t}b,p^ka)$,
i.e. $0=p^kac+p^{k+t}b$ and $p^k(c+p)=p^ka$, which are equivalent to
that $p^k(c+p)c=-p^{k+t}b\in p^{k+t}A_f$, i.e. $\|(c+p)c\|_p\geq t$, where $c\in p(A_f/\langle p^t\rangle)$, $1\leq t\leq s-k-1$ and $1\leq k\leq s-2$. By $1\leq k\leq s-2$, it follows that $s\geq 3$.

\par
   When $s=3$, we have $k=1$ and $t=1$. In this case, an argument similar to (viii-1) shows that $c=0$.

\par
  In the following, we assume that $s\geq 4$.
Then we have one of the following three subcases:

\par
   (ix-1) When $t=1$, then $c\in p(A_f/\langle p\rangle)=\{0\}$, i.e., $c=0$. In this case, $\|(c+p)c\|_p=\|0\|_p=s>t$.

\par
   (ix-2) When $t\geq 2$, from $1\leq t\leq s-k-1$ we deduce that $1\leq k\leq s-3$. By $c\in p(A/\langle p^t\rangle)$ and Lemma 2.1(i) it follows that
$c=e_1p+\ldots+e_{t-1}p^{t-1}$ where $e_1,\ldots,e_{t-1}\in T_f$.
Then $c+p=(e_1+1)p+\ldots+e_{t-1}p^{t-1}$. If $t=2$, it is obvious that $\|(c+p)c\|_p\geq 2=t$ for all
$c\in p(A/\langle p^2\rangle)=\{e_1p\mid e_1\in T_f\}$.

\par
   Now, assume $t\geq 3$. Then $1\leq k\leq s-4$ and $s\geq 5$.
Suppose that $e_1\neq 0$ and $e_1\neq -1$. Then $\|c\|_p=\|c+p\|_p=1$ and $\|(c+p)c\|_p=2<t$. Hence
By $\|(c+p)c\|_p\geq t\geq 3$, we have
one of the following two subcases:

\par
  $\triangleright$ $e_1\neq 0$ and $e_1=-1$. In this case, $\|c\|_p=1$. Then $\|(c+p)c\|_p\geq t$ if and only if
$\|c+p\|_p\geq t-1$, which is equivalent to that $e_1=-1$ and $c+p=e_{t-1}p^{t-1}$, i.e. $c=-p+e_{t-1}p^{t-1}$,
where $e_{t-1}\in T_f$.

\par
  $\triangleright$ $e_1\neq -1$ and $e_1=0$. In this case, $\|c+p\|_p=1$. Then $\|(c+p)c\|_p\geq t$ if and only if
$\|c\|_p\geq t-1$, which is equivalent to that $c=e_{t-1}p^{t-1}$ where $e_{t-1}\in T_f$.



\end{document}